\documentclass{emulateapj}

\countdef\decade=200
\advance\decade by \year
\countdef\hours=201
\advance\hours by \time
\divide\hours by 60
\countdef\mins=202
\advance\mins by \hours
\multiply\mins by 60
\multiply\hours by 100
\countdef\miltime=203
\advance\miltime by \hours
\advance\miltime by \time
\advance\miltime by -\mins

\shorttitle{Massive Core Collapse and Fragmentation}
\shortauthors{Krumholz, Klein, \& McKee}

\newcommand{\ltsim}{\protect\raisebox{-0.5ex}{$\:\stackrel{\textstyle <}
        {\sim}\:$}}
\newcommand{\gtsim}{\protect\raisebox{-0.5ex}{$\:\stackrel{\textstyle >}
        {\sim}\:$}}

\newcommand{\calp}{\mathcal{P}}
\newcommand{\kr}{\kappa_{\rm R}}
\newcommand{\kp}{\kappa_{\rm P}}

\newcommand{\vecv}{\mathbf{v}}
\newcommand{\vecx}{\mathbf{x}}

\newcommand{\vecF}{\mathbf{F}}
\newcommand{\vecn}{\mathbf{n}}

\newcommand{\msun}{M_{\odot}}
\newcommand{\lsun}{L_{\odot}}

\begin{document}

\title{Radiation-Hydrodynamic Simulations of Collapse and
Fragmentation in Massive Protostellar Cores}

\slugcomment{Accepted for publication in the Astrophysical Journal, November 4, 2006}

\author{Mark R. Krumholz\footnote{Hubble Fellow}}
\affil{Department of Astrophysical Sciences, Princeton University,
Princeton, NJ 08544}
\email{krumholz@astro.princeton.edu}

\author{Richard I. Klein}
\affil{Astronomy Department, University of California, Berkeley,
Berkeley, CA 94720, and Lawrence Livermore National Laboratory,
P.O. Box 808, L-23, Livermore, CA 94550}
\email{klein@astron.berkeley.edu}

\author{Christopher F. McKee}
\affil{Departments of Physics and Astronomy, University of California,
Berkeley, Berkeley, CA 94720}
\email{cmckee@astron.berkeley.edu}

\begin{abstract}
We simulate the early stages of the evolution of turbulent,
virialized, high-mass protostellar cores, with primary attention to
how cores fragment, and whether they form a small or large number
of protostars. Our simulations use the Orion adaptive mesh refinement
code to follow the collapse from $\sim 0.1$ pc scales to $\sim 10$ AU
scales, for durations that cover the main fragmentation phase, using
three-dimensional gravito-radiation hydrodynamics. We find that for a
wide range of initial conditions radiation feedback from accreting
protostars inhibits the formation of fragments, so that the vast
majority of the collapsed mass accretes onto one or a few
objects. Most of the fragmentation that does occur takes place in
massive, self-shielding disks. These are driven to gravitational
instability by rapid accretion, producing rapid mass and angular
momentum transport that allows most of the gas
to accrete onto the central star rather than forming fragments.
In contrast, a control run using the same initial conditions but an
isothermal equation of state produces much more fragmentation, both in
and out of the disk. We
conclude that massive cores with observed properties are not likely to
fragment into many stars, so that, at least at high masses, the core
mass function probably determines the stellar initial mass
function. Our results also demonstrate that simulations of massive
star forming regions that do not include radiative transfer, and
instead rely on a barotropic equation of state or optically thin
heating and cooling curves, are likely to produce misleading results.
\end{abstract}


\section{Introduction}

The previous generation of telescopes revealed a great deal about
the gas from which massive stellar clusters form. With them, observers
were able to survey the dense clumps of thousands of $\msun$ that are
likely the progenitors of clusters, using molecular line emission
\citep[e.g.][]{plume97,shirley03}, thermal dust emission
\citep[e.g.][]{carey00,mueller02}, or infrared absorption
\citep[e.g.][]{menten05,rathborne05,rathborne06a,simon06}. However,
the large distances to these regions, their high extinctions, and the
confusion produced by their density prevented these observations from
directly probing structures with masses comparable to individual
stars, data that have been available since the 1980s for nearby, low-mass
star-forming regions. In the last few years, millimeter
interferometers and the Spitzer Space Telescope have started to
change that situation by
making available information on the structure of massive star forming
regions comparable to that previously available only for low-mass
regions. These observations have identified a population of high-mass
cores that are comparable in mass to individual massive
stars. They are dense (mean densities $\sim 10^6$ H nuclei cm$^{-3}$,
rising strongly towards their centers), cold (temperatures $\approx
10-40$ K), turbulent (linewidths $\sim 1$ km s$^{-1}$),
compact (radii $\sim 0.1$ pc), and round (aspect ratios of $2:1$ or
less) \citep{reid05,sridharan05,beuther05b,beuther06a,garay05,
pillai06a}. In some cases they show no mid-infrared emission or even
mid-infrared absorption, indicating that they have not yet converted a
significant fraction of their mass into stars, and are therefore near
the onset of star formation.

The idea that these cores might be the progenitors of individual
massive stars is bolstered by two pieces of circumstantial
evidence. First, the mass function of these cores appears to match the
\citet{salpeter55} slope of roughly $-1.3$ in the logarithmic
distribution observed for high-mass end of the stellar initial mass
function \citep[IMF,][]{beuther04b, reid05, reid06a, reid06b}. This
extends earlier observations showing
that in nearby low-mass star-forming regions the core mass function
matches the IMF as well \citep{motte98,testi98,johnstone01,
onishi02}. Second, cores are mass segregated in such a manner that the
mass function is the same throughout a protocluster gas clump, with
the exception that the most massive cores, those greater than several
$\msun$ in mass, are found only near the center
\citep{elmegreen01,stanke06}. Star clusters exhibit a very similar
pattern of mass segregation \citep{hillenbrand98,huff06}, and while
some of this may be dynamically produced, much of it is likely a result
of the locations where the stars formed \citep{bonnell98b,tan06a}.

However, a direct mapping from core mass to star mass is only
possible if massive cores collapse to form individual stars or
small-multiple systems, as proposed by \citet[hereafter MT02 and
MT03]{mckee02, mckee03}, rather than fragmenting into many
objects and producing a cluster of low-mass stars. Whether this
happens or not is quite uncertain. \citet{bate05} argue that dense
cores are likely to produce small objects because the Jeans mass
decreases with density at fixed temperature \citep[although
see][]{martel06}. \citet{dobbs05} simulate the collapse of massive,
turbulent cores and find that they generally fragment into as many as 20
objects, depending on initial conditions and on the assumed gas
equation of state. However, \citet{krumholz06b} uses one dimensional
analytic calculations to show that radiation feedback
from accreting protostars can substantially inhibit fragmentation even
at early times, because at the high accretion rates and opacities
expected in massive cores, accretion luminosity can heat gas to
hundreds of Kelvin out to distances of $\gtsim 1000$ AU from an
accreting protostar. \citeauthor{krumholz06b} also finds that using an
isothermal or barotropic equation of state, as \citet{bate05} and
\citet{dobbs05} do, is likely to produce misleading results on
fragmentation because it misses this effect.

While these calculations are suggestive, because they are analytic
they are necessarily limited in how they deal with real, turbulent
cores. The best means of settling the question of how massive cores
fragment is direct numerical simulation, including a treatment of
radiative feedback from embedded protostars. However, such simulations
have to this point not been reported in the literature. Some
simulations of massive star formation with radiation use
quiescent initial conditions \citep{yorke99, yorke02} in
two dimensions, and are therefore incapable of answering questions
about the fragmentation of turbulent structures. (Those calculations
focus on the effects of radiation pressure, an important effect in the later
evolution of massive protostars which we do not consider in detail in
this paper.) In three dimensions, some simulations of massive cores
use local cooling functions rather than solving the radiative transfer
problem \citep{banerjee06}, and are therefore
unable to study the effects of feedback. Moreover, the modifications
to the cooling function the authors use to approximate the behavior of
optically thick gas are of unknown accuracy. Simulations of star
formation with feedback and a treatment of radiative transfer
have been limited to low-mass, non-turbulent initial conditions
\citep{whitehouse06}. Furthermore, both \citeauthor{whitehouse06} and
\citeauthor{banerjee06}\ only advance their
simulations to the point where the first collapsed object forms, and
for this reason they are incapable of studying accretion and
fragmentation, or the effects of radiative feedback on either of these
processes.

Here we report the first three-dimensional gravito-radiation
hydrodynamic simulations of the collapse and fragmentation of
turbulent high-mass protostellar cores. Our simulations include
radiative transfer and the effects of feedback from both accretion
onto and nuclear burning within embedded protostars. We run through
the main
fragmentation phase and follow the accretion process to the point
where deuterium burning begins in the most massive stars, thereby
greatly heating the gas and strongly suppressing further
fragmentation. This enables us to address the question of how
fragmentation of massive cores proceeds, and how radiative feedback
influences it. In \S~\ref{method} we discuss the methodology for our
simulations, and in \S~\ref{results} we present our results. We
discuss the implications of these results for the mechanism of massive
star formation and the origin in of the IMF in \S~\ref{discussion},
and summarize in \S~\ref{conclusion}.

\section{Simulation Methodology}
\label{method}

\subsection{Evolution Equations}

We describe the evolution of a massive protostellar core using the
equations of gravito-radiation hydrodynamics in the thermal radiation
flux-limited diffusion approximation. Written in conservation form,
\citet{krumholz07b} show that these equations to leading order in
$v/c$ are
\begin{eqnarray}
\label{massconservation}
\frac{\partial \rho}{\partial t} + \nabla\cdot (\rho \vecv) & = & 0, \\
\frac{\partial}{\partial t} (\rho \vecv) + \nabla \cdot (\rho \vecv \vecv) 
& = & \qquad\qquad\qquad\qquad\qquad 
\nonumber \\
& & 
\!\!\!\!\!\!\!\!\!\!\!\!\!\!\!\!\!\!\!\!\!\!\!\!
\!\!\!\!\!\!\!\!\!\!\!\!\!\!\!\!\!\!\!\!
\lefteqn{
{} - \nabla P - \rho \nabla \phi - \lambda \nabla E
}
\label{momentumconservation} \\
\frac{\partial}{\partial t} (\rho e) + \nabla \cdot \left[\left(\rho e +
P\right) \vecv\right] & = & 
\nonumber \\
& & 
\!\!\!\!\!\!\!\!\!\!\!\!\!\!\!\!\!\!\!\!\!\!\!\!
\!\!\!\!\!\!\!\!\!\!\!\!\!\!\!\!\!\!\!\!
\lefteqn{-\rho \vecv \cdot \nabla \phi 
- \kp \rho (4\pi B - cE) + \lambda \vecv \cdot \nabla E}
\label{energyconservation} \\
\frac{\partial E}{\partial t} - \nabla \cdot
\left(\frac{c\lambda}{\kr}\nabla E\right)
& = &
\nonumber \\
& &
\!\!\!\!\!\!\!\!\!\!\!\!\!\!\!\!\!\!\!\!\!\!\!\!
\!\!\!\!\!\!\!\!\!\!\!\!\!\!\!\!\!\!\!\!
\lefteqn{
\sum_i L_i \delta(\vecx - \vecx_i) + \kp \rho (4\pi B - c E) }
\nonumber \\
& &
\!\!\!\!\!\!\!\!\!\!\!\!\!\!\!\!\!\!\!\!\!\!\!\!
\!\!\!\!\!\!\!\!\!\!\!\!\!\!\!\!\!\!\!\!
\lefteqn{
{} - \lambda \vecv \cdot \nabla E 
}
\nonumber \\
& &
\!\!\!\!\!\!\!\!\!\!\!\!\!\!\!\!\!\!\!\!\!\!\!\!
\!\!\!\!\!\!\!\!\!\!\!\!\!\!\!\!\!\!\!\!
\lefteqn{
{} - \nabla \cdot
\left[\frac{3-R_2}{2} E \vecv + 
\frac{3 R_2 - 1}{2} E \vecv \cdot (\vecn \vecn)\right].
}
\label{radenergy}
\end{eqnarray}
Here $\rho$, $\vecv$, $e$, and $P$ are the density, velocity,
non-gravitational specific energy (thermal plus kinetic), and thermal
pressure of the gas, $\phi$ is the gravitational potential, $E$ is the
radiation energy density, $B = c a_R T_g^4/(4\pi)$ is the Planck
function of the gas temperature $T_g$, $\kp$ is the
Planck-mean specific opacity of the gas measured in its rest frame,
$\lambda$ and $R_2$ are dimensionless numbers describing the radiation
field whose significance we discuss below, $L_i$ and $\vecx_i$ are the
luminosity and position of the $i$th star, and $\vecn$ is a unit
vector anti-parallel to $\nabla E$. To leading order in $v/c$, these
equations match those of other flux-limited diffusion
radiation-hydrodynamic codes, e.g. ZEUS \citep{hayes06}.

The gas pressure, specific energy, and temperature $T_g$ are related
by an ideal equation of state
\begin{equation}
P = (\gamma-1) \left(e-\frac{1}{2}\rho v^2\right)
= \frac{\rho k_B T_g}{\mu},
\end{equation}
where $\mu=2.33 m_H$ is the mean particle mass in a gas of molecular
hydrogen and helium with the standard cosmic abundance, and we
approximate $\gamma=5/3$ since over most of the volume the gas is too
cool to excite rotational or vibrational modes of hydrogen. In
practice the choice of $\gamma$ has almost no effect, because
radiative time scales are generally shorter than mechanical ones, so
the gas temperature and therefore the effective equation of state is
essentially fixed by radiative transfer effects.

The gravitational potential is determined by Poisson's equation,
including the contribution from stars, which we treat as point
masses:
\begin{equation}
\label{poisson}
\nabla^2 \phi = 4\pi G \left[\rho + \sum_i M_i
\delta(\vecx-\vecx_i)\right],
\end{equation}
where $M_i$ is the mass of the $i$th star.

The dimensionless numbers appearing in radiation-related terms are the
flux limiter $\lambda$ and the Eddington factor $R_2$. Their purpose
is to interpolate between the optically thick and optically thin
limits. They are defined by
\begin{eqnarray}
\lambda & = & \frac{1}{R} \left(\coth R - \frac{1}{R}\right), \\
R & = & \frac{|\nabla E|}{\kr \rho E}, \\
R_2 & = & \lambda + \lambda^2 R^2,
\end{eqnarray}
where $\kr$ is the Rosseland-mean specific opacity of the gas.
The flux limiter has the property that $\lambda\rightarrow 1/3$ in
optically thick regions and $\lambda \rightarrow \kr \rho E/|\nabla E|$
in optically thin regions. For optically thick flows this behavior
means that the flux in the frame comoving with the gas approaches
$\vecF \rightarrow-c/(3\kr\rho)\nabla E$, the correct value for diffusion.
For optically thin flows it limits to $\vecF\rightarrow c E
\vecn$, so that the effective propogation speed of the
radiation is limited to $c$. Similarly, for optically thick regions
$R_2\rightarrow 1/3$, which sets the comoving-frame radiation pressure
tensor to the correct isotropic behavior, $\calp \rightarrow
(E/3)\mathbf{I}$, where $\mathbf{I}$ is the identity tensor. For
optically thin flows $R_2\rightarrow 1$, which gives $\calp\rightarrow
E \vecn\vecn$, the correct limiting value for free-streaming
radiation. We refer readers to \citet{krumholz07b} for a detailed
treatment of the relationship between the comoving frame and lab frame
quantities, and how the values of $\lambda$ and $R_2$ are related to
comoving frame quantities.

Our equations are easy to understand intuitively. The term
$-\lambda\nabla E$ in the momentum equation
(\ref{momentumconservation}) simply represents the
radiation force $\kr\rho\vecF/c$, neglecting distinctions between the
comoving and laboratory frames which are smaller than order
$v/c$. Similarly, the terms $-\kp\rho(4\pi B - cE)$ and
$\lambda\vecv\cdot\nabla E$ in the gas energy equation
(\ref{energyconservation}) represent radiation absorbed minus
radiation emitted by the gas, and the work done by the
radiation field on the gas. In the radiation energy equation
(\ref{radenergy}), the second term on the left hand side is the
divergence of the radiation flux, i.e. the rate at which radiation
diffuses, and the terms on the right hand side describe, from left to
right, radiation emitted by protostars, radiation emitted minus
radiation absorbed by gas, work done by the gas on the radiation
field, and advection of radiation enthalpy by the gas.

Note that our equations correspond to those of \citet{krumholz07b}
for the static diffusion case, which \citeauthor{krumholz07b}\ show is
the relevant limit for massive protostellar envelopes, with two
differences. First is the addition of the terms describing gravity and
point sources of radiation, which \citeauthor{krumholz07b}\
do not include. Second is a difference in the coefficient of the work
term, $\lambda\vecv\cdot\nabla E$. Due to this difference, the
equations we give here are only accurate to leading order in $v/c$,
rather than to first order. In practice, this should make little
difference in the outcome, since $v/c\ll 1$ everywhere in our
calculation.

We discuss the applicability of our equations, including the
limitations imposed by the approximations we adopt, in
\S~\ref{limitations}.

\subsection{Models for Dust and Protostars}
\label{models}

To complete our specification of the problem, we must adopt models to
describe the dust, the primary source of opacity, and protostellar
evolution. We approximate that the dust and gas are well-coupled, and
neglect the possibility that the grain population evolves with time or
with position except as a function of local gas properties. We also
assume that dust grains react quickly to changes in temperature, so
that changes in opacity due to changes in the grain population (such
as sublimation of certain grain components) occur without any time
delay. This enables us to specify the Planck- and Rosseland-mean
opacities as simple functions of the radiation temperature. We adopt the
dust model of \citet{pollack94}, which includes six species of dust
grains, each with its own sublimation temperature. We approximate the
tabulated opacities computed by \citeauthor{pollack94} using a simple
piecewise-linear analytic formula. Our fit gives
\begin{equation}
\kp =
\left\{
\begin{array}{lr}
0.3 + 7.0\,(T_r/375)
, \qquad &
T_r \leq 375 \\
7.3 + 0.7\,(T_r - 375)/200
, &
375 < T_r \leq 575 \\
3.0 + 0.1\,(T_r - 575)/100
, &
575 < T_r \leq 675 \\
2.8 + 0.3\,(T_r - 675)/285
, &
675 < T_r \leq 960 \\
3.1 - 3.0\,(T_r - 960)/140
, &
960 < T_r \leq 1100 \\
0.1, &
T_r > 1100
\end{array}
\right.,
\end{equation}
and
\begin{equation}
\kr =
\left\{
\begin{array}{lr}
0.1 + 4.4\,T_r/350
, \qquad &
T_r \leq 350 \\
3.9
, &
350 < T_r \leq 600 \\
0.7
, &
600 < T_r \leq 700 \\
0.25
, &
700 < T_r \leq 950 \\
0.25 - 0.15\,(T_r - 950)/50
, &
950 < T_r \leq 1000 \\
0.1, &
T_r > 1000
\end{array}
\right.,
\end{equation}
where the radiation temperature $T_r$ is in units of K and $\kp$ and
$\kr$ are in units of cm$^2$ g$^{-1}$. Note that, due to the optical
thickness of a massive core, $T_r\approx T_g$ everywhere within one
except near its surface. Also note that at high temperatures where the
dust has sublimed, our choice to set $\kp=\kr=0.1$ cm$^2$ g$^{-1}$ is
purely a numerical convenience we use to represent a ``small''
opacity. The true opacity depends in detail on the radiation spectrum
and the physical state of the gas (molecular, atomic, or ionized), but
is certainly much smaller than the opacity due to dust
grains. However, sharp opacity gradients make it difficult for our
radiation iterative solver to converge, so the choice of 0.1 cm$^2$
g$^{-1}$ is a compromise between physical realism and numerical
efficiency. This choice has little effect in practice, because for the
simulations we describe here only a handful of computational cells
reach temperatures high enough to be in this regime.

\begin{figure}
\plotone{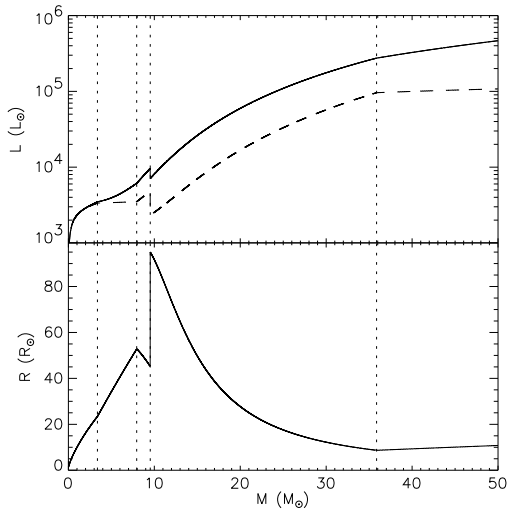}
\caption{
\label{starmodel}
Luminosity (\textit{upper panel}) and radius (\textit{lower panel}) of
a protostar versus mass computed using our protostellar
model with a constant accretion rate of $10^{-3}$
$\msun$ yr$^{-1}$. The dashed line in the upper panel is the
luminosity due to accretion. The dotted vertical lines mark, from
left to right, the masses at which deuterium burning starts, deuterium
in the core is exhausted, convection in the envelope starts, and
hydrogen burning starts.\\
}
\end{figure}

The final piece of our physical model is a method to specify the
luminosity of accreting protostars, which appears as a source term in
the radiation energy equation. The input to this model is the mass
accretion history of the protostar, which is determined with the sink
particle algorithm of \citet{krumholz04}, which we discuss in more
detail in \S~\ref{algorithm}. We adopt the protostellar evolution
model of MT03, an extension of earlier
models by \citet{nakano95,nakano00}. The model is fairly complex, so
we refer readers to MT03 for a detailed description, but we
summarize its central features here. The model describes a star as a
polytropic sphere, and computes the evolution of the protostellar
radius, central temperature, luminosity, and polytropic index using
the equation of conservation of energy for the star, including terms
that describe the energy used to dissociate and ionize the incoming
gas and the energy released by deuterium and hydrogen burning. The
model includes approximate treatments of the onset of deuterium burning
in the core, the exhaustion of deuterium in the core, the formation of
a radiative barrier and the formation to a convective envelope, and
the start of hydrogen burning. It reproduces the detailed numerical
simulations of \citet{stahler88} and \citet{palla92} to $\sim
10\%$. Figure \ref{starmodel} shows a sample calculation using our
numerical implementation of the model for the case of a protostar
accreting at a constant rate of $10^{-3}$ $\msun$ yr$^{-1}$.

\subsection{Solution Algorithm}
\label{algorithm}

We solve the evolution equations using the Orion
adaptive mesh refinement (AMR) code. The code consists of three
primary modules, which operate sequentially in each update
cycle. 

First is the hydrodynamics module \citep{puckett92,
truelove98, klein99}, which solves the Euler equations of gas dynamics
(\ref{massconservation}), (\ref{momentumconservation}), and
(\ref{energyconservation}) including all of
the terms except those involving radiation. The
hydrodynamics module uses a conservative Godunov scheme with an
approximate optimized Riemann solver \citep{toro97}. The algorithm is
second-order accurate in time and space for smooth flows, and provides
a robust treatment of shocks and discontinuities using very little
artificial viscosity.

The second part of the code is the gravity module, which computes the
gravitational potential from the Poisson equation (\ref{poisson})
using a multigrid iteration scheme \citep{truelove98, klein99, fisher02}.

The third part is a radiation module, which is updated using the
\citet{krumholz07b} operator splitting approach, in which we update
the dominant radiation terms describing diffusion and emission minus
absorption implicitly using the approach of \citet{howell03}, while we
update the work and advection terms explicitly. This algorithm is
stable and accurate for problems in the static diffusion limit such as
ours, and we refer readers to \citet{howell03} and \citet{krumholz07b}
for a detailed discussion of the algorithm and the tests we have
performed with it. For the radiation update, we use the dust model
described in \S~\ref{models} to compute the opacities, and we use the
luminosity of each protostar, computed as described below, as a source
term.

We supplement these modules by using the Eulerian sink particle
algorithm of \citet{krumholz04} to handle the formation of
protostars. When a cell on the finest AMR level violates the Jeans
condition for gravitational stability \citep{truelove97}, we create a
``star particle'' in that cell. Each such particle is a sink particle,
as described by \citet{krumholz04}, but also has a corresponding
protostellar model, which in addition to the mass includes the star's
radius, luminosity, polytropic index, accretion rate, mass of
deuterium remaining, and phase of evolution (e.g. whether the star has
developed a convective envelope yet). In each update cycle, after
completing the standard update step for sink particles, we also update
the protostellar model. When we perform a radiation update, the protostellar
luminosity becomes a source term in the radiation energy equation.

All of these pieces operate with the AMR framework \citep{berger84,
berger89, bell94}. We cover the computational domain with a series of
levels $l=0,1,2,\ldots L$, where $l=0$ is the coarsest level, which
covers the entire computational domain. Each level is a union of
rectangular grids, which need not be contiguous. The grids are
nested, such that every grid on level $l>0$ is entirely enclosed
within one or more grids on level $l-1$. Grids on a given level all
have the same grid spacing $\Delta x^l$, and spacings on different
levels are related by integer ratios $f>1$, so that $\Delta x^{l+1} =
\Delta x^l/f$. For all the calculations we present here, we use
$f=2$. Each level advances with its own time step $\Delta t^l$, and
time steps on adjacent levels
obey the relation $\Delta t^{l+1} = \Delta t^l/f$. The process for
advancing the calculation is recursive. To advance a time step on
level $l$, one first updates all the level $l$ cells through a time
$\Delta t^l$, then updates all the cells on level $l+1$ through $f$
time steps of size $\Delta t^{l+1}$. However, after
completing each level $l+1$ update, one advances the cells on level
$l+2$ through $f$ time steps, and so forth down to the finest level
present. After every $f$ cycles on each level $l>0$, we perform a
synchronization procedure between levels $l$ and $l-1$ to ensure that
mass, momentum, and energy are conserved across level boundaries.

The overall time step is set by the Courant condition computed on each
level,
\begin{equation}
\Delta t^l = C \frac{\Delta x^l}{\max(|\vecv| + c_{\rm eff})},
\end{equation}
where the maximum is taken over all cells on that level. The effective
sound speed is
\begin{equation}
c_{\rm eff} = \sqrt{\frac{\gamma P + (4/9) E 
\left(1-e^{-\kr \rho \Delta x}\right)}{\rho}},
\end{equation}
where $\gamma$ is the ratio of specific heats for the gas, and the
factor $(1-e^{-\kr \rho \Delta x})$ provides a means of interpolating between
optically thick cells, where radiation pressure contributes to the
restoring force and thus increases the effective signal speed, and
optically thin cells, where radiation does not provide any
pressure. To enforce the integer ratio between
time steps on different levels, after computing the time step $\Delta
t^l$ on each level, we set the time step on level 0 to $\Delta t^0 =
\min(\Delta t^l f^l)$, then reset the time step on all other levels to
$\Delta t^0/f^l$.

\subsection{Initial, Boundary, and Refinement Conditions}

\begin{figure}
\plotone{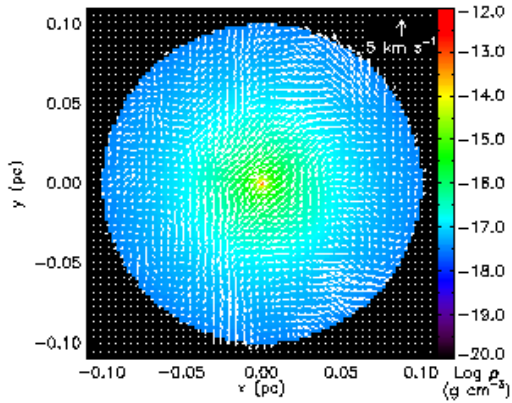}
\caption{
A slice through the $xy$ plane showing the density
(\textit{grayscale}) and velocity (\textit{arrows}) fields at the
start of run 100A.
\label{100Ainit}
}
\end{figure}

We chose initial conditions that correspond to the analytic
model of MT03 for high-mass cores. Each calculation begins
with an initial core that is a sphere of gas of mass $M$ and radius
$r_1$. The core is centrally concentrated, with density profile
$\rho\propto r^{-3/2}$, down to some inner radius $r_0$. Inside $r_0$
the density is constant. This corresponds approximately to the
thermally-supported core and turbulently-supported envelope of a
MT03 core. To impose the initial turbulent velocity
field, we generate a $1024^3$ grid of perturbations with a power
spectrum $P(k)\propto k^{-2}$ using the method of \citet{dubinski95},
corresponding to the spectrum expected for supersonic, shock-dominated
turbulence. We overlay the perturbation cube on the core such that the
core just fits inside the perturbation cube, and assign the velocity
at every point in the cube to the corresponding point inside the
core. The use of a $1024^3$ grid allows turbulent power to be present
down to scales of $1/1024$ of a cloud diameter, which is approximately
the diameter of the non-turbulent, thermally supported region in the
MT03 model. We scale the total velocity of the core so that
the one-dimensional core velocity dispersion is $\sigma =
(GM/r_1)^{1/2}$, the dispersion required for the core to be in
approximate hydrostatic balance. We do not drive the turbulence or
otherwise inject energy after the simulation starts.  We set the
initial temperature of the core to $T_g=20$ K, and the initial
radiation energy density to $1.21\times 10^{-9}$ erg cm$^{-3}$,
the energy density of a blackbody radiation field at $T_r=20$ K
temperature.

Outside the core is an ambient medium with density 100 times smaller
than the density at the edge of the core and temperature 100 times
higher, so the core and ambient medium are in thermal pressure
balance. The opacity of the ambient medium is set to zero, so that it
cannot cool, radiatively heat the core, or inhibit the escape of
radiation from the core.

We place the core and ambient medium inside a computational cube
centered on the origin with length $L$, chosen large enough so that no
core material ever approaches the edge of the computational domain. At
the boundary we impose symmetry conditions on the hydrodynamic
evolution, Dirichlet conditions for the graviational field, with the
potential on the boundary set equal to $-GM/r$, and Marshak boundary
conditions on the radiation field. Marshak conditions are a variant of
Neumann conditions in which the flux into the computational domain is
set to a constant value of $cE_0/2$, where we set $E_0$ equal
to the initial radiation energy density. The flux out of the
computational domain at the face of each cell on the boundary is set
equal to $cE/2$, where $E$ is the radiation energy density in the
cell. This condition imposes a uniform 20 K radiation bath, but allows
excess radiation generated inside the computational domain to escape
freely.

The final piece of our computational setup is the refinement
conditions, which determine when new
high resolution grids are created or when existing ones are removed,
a process that occurs automatically throughout the calculation.
We use three refinement criteria. First, in order to prevent numerical
fragmentation we refine any cell on level $l$ in which the density of
that cell violates the Jeans condition for self-gravitational
stability \citep{truelove97}, 
\begin{equation}
\rho < \rho_J = J^2 \frac{\pi c_s^2}{G (\Delta x^l)^2}.
\end{equation}
We use a Jeans number $J=1/8$. Second, we refine any cell whose
distance from the nearest sink particle is less than $16\Delta
x^l$, so that the region around sink particles is always
well-resolved. Third, to ensure that we do not produce artificially
large radiation pressure forces due to poorly resolved radiation
energy density gradients, we refine any cell in which $\Delta x^l
|\nabla E|/E > 0.25$, i.e. anywhere the radiation energy density
changes by more than 25\% per cell. All of these refinement criteria
are applied up to a maximum level $L_{\rm max}$, which we specify when
we begin the calculation. All runs use a resolution of $128^3$ cells
on level 0 and a maxmimum level of 7, giving an effective resolution
of $16384^3$.

We perform four runs, whose properties we summarize in Table
\ref{runsetup}. Run 100A is our baseline run to which we compare the
others. We show the initial density and velocity field for this run in
Figure \ref{100Ainit}. Run 100B uses the same parameters as run 100A
and has a turbulent velocity field with the same power spectrum, but
a different random realization than run 100A. This allows us to study
how the random velocity field influences the results. Run 200A uses
the same velocity field as 100A, but a more massive core, enabling us
to study how our results depend on initial core mass. We keep the mean
column density constant, so that we are not changing the average
opacity to radiation. Finally, run 100ISO uses the
same initial conditions as run 100A, but for it we do not use
radiative transfer. Instead, we use an isothermal equation of state
fixed to $T_g=20$ K. This amount to setting $E=0$ and $B=0$ in
equations (\ref{momentumconservation}) and (\ref{energyconservation}),
and dropping equation (\ref{radenergy}) entirely. This lets us isolate
how radiative transfer affects the results, and study whether
simulations that do not include it are reliable. Run 100ISO is very
similar in setup to some of the models evaluated by \citet{dobbs05}.

\begin{deluxetable*}{cccccccccccc}
\tablecaption{Simulation parameters\label{runsetup}}
\tablehead{
\colhead{Run name} &
\colhead{$M$ $(\msun)$} &
\colhead{Field} &
\colhead{EOS} &
\colhead{$r_1$ (pc)} &
\colhead{$r_0$ (AU)} &
\colhead{$L$ (pc)} &
\colhead{$\Delta x^{L_{\rm max}}$ (AU)} &
\colhead{$\rho_1$ ($10^{-14}$ g cm$^{-3}$)} &
\colhead{$t_{\rm ff}$ (kyr)} &
\colhead{$\sigma$ (km s$^{-1}$)}
}
\startdata
100A & 100 & A & RT & 0.1 & 38.4 & 0.6 & 7.5 & 1.0 & 52.5 & 1.7\\
100B & 100 & B & RT & 0.1 & 38.4 & 0.6 & 7.5 & 1.0 & 52.5 & 1.7\\
200A & 200 & A & RT & 0.14 & 53.5 & 0.85 & 10.7 & 0.72 & 62.4 & 2.0\\
100ISO & 100 & A & ISO & 0.1 & 38.4 & 0.6 & 7.5 & 1.0 & 52.5 & 1.7\\
\enddata
\tablecomments{Col. (3): Perturbation field, A or B. Col. (4):
Equation of state, RT = radiative transfer, ISO = isothermal. Col. (7)
Grid spacing on finest AMR level. Col (8): Initial density of inner,
constant density region. Col (9): Free-fall time at the mean density.
}
\end{deluxetable*}

\section{Results}
\label{results}

We evolve each run for 20 kyr, which is 38\% of a mean-density
free-fall time for runs 100A, 100B, and 100ISO, and 32\% of a
mean-density free-fall time for run 200A. For the radiative runs, in
all cases the most massive star formed by the end of the run has begun
deuterium burning, and thus should rise rapidly in luminosity
thereafter (see Figure \ref{starmodel}), strongly inhibiting further
fragmentation. The portion of the evolution we follow therefore
includes the primary
fragmentation phase, during which the densest parts of the core
collapse and the pattern of fragmentation is established. As we
discuss in \S~\ref{diskfragment}, there may be subsequent secondary
fragmentation in unstable protostellar disks at later times, but this
does not significantly change where most of the collapsing mass goes.

In the analysis that follows, we only consider sink particles stars if
they have a mass of $0.05$ $\msun$ or more, the mass at which 
``second collapse'' to protostellar densities occurs
\citep{masunaga98, masunaga00}. (Smaller mass objects can still
collapse to form stars or brown dwarfs, but they produce insufficient
pressure to produce rapid dissociation of molecular hydrogen, leading
to second collapse. Instead, they contract much more slowly.)
We take this precaution because a few times
over the course of a run our radiation implicit solver fails to
converge and produces an unrealistically low temperature in an isolated
cell. The low temperature cell may be Jeans unstable and form a sink
particle (see \citealt{krumholz04} for a discussion of the sink
particle creation algorithm). These artificially-created sink
particles are harmless because they contain negligible mass -- at most
a few times $10^{-2}$ $\msun$, usually much smaller. 
Since the radiation solver generally recovers on the next time step
and produces a reasonable temperature, and the regions around the sink
particles are gravitationally stable once a normal temperature is
restored, these low-mass sink particles do not accrete or radiate at a
noticable rate, and never contain a significant amount of mass. We
impose a mass threshold before we consider a sink particle to be a
protostar in order to eliminate these spurious objects. This condition
may cause us to mischaracterize some real, non-numerical but very low
mass stars, but the amount of mass in stars we could miss in this
fashion is obviously tiny. Although the
isothermal run is clearly not subject to this problem, we impose the
same condition on our analysis of it to avoid introducing any
bias.

Each of the radiative runs required roughly $60,000-70,000$ CPU-hours
on an IBM SP, running in parallel on 128 processors. The isothermal
run required approximately $10,000$ CPU-hours.

\subsection{Summary of Evolution}

We summarize the evolution of our runs starting with run 100A, and we
then discuss how the other runs differ. We defer discussion of how
fragmentation and protostar formation occurs across the runs until
\S~\ref{fragstats}, and here focus on the overall qualitative
morphology of the gas. 

\subsubsection{Run 100A}

\begin{figure*}
\plotone{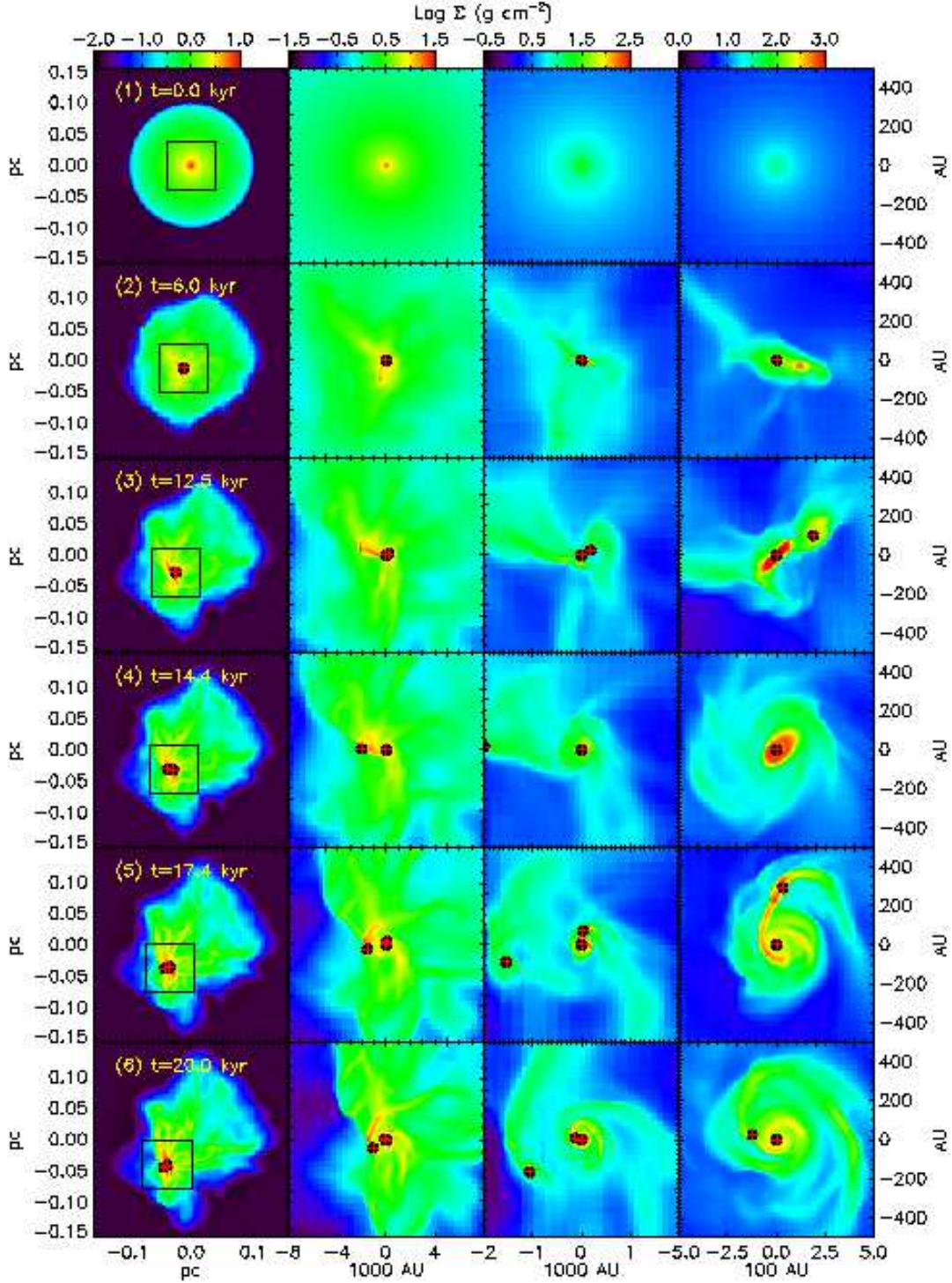}
\caption{
\label{100Aevol}
Column density as a function of time in run 100A. From top to bottom,
the rows show the cloud state at increasing time, as indicated. From
left to right, each step to the right corresponds to decreasing the
linear size of the region displayed by a factor of 4, from a $0.31$ pc
region in the left column to a $1000$ AU region in the right
column. At the top of each column we give a scale bar for the images
in that column. In the left column the region shown is always
centered on the origin, and the region shown in the second column is
indicated by the black box; in the other columns, the region shown is
centered on the location of the primary star at that time. Stars are
indicated by the white plus signs. All images are shown in the same
projection.\\
}
\end{figure*}

Figure \ref{100Aevol} shows a
time-sequence of the evolution of the run, starting from the initial
state shown in row (1). Turbulent motions delay the onset of
collapse for a while, but as the turbulence decays gas starts to
collapse. The first object, which we refer to hereafter as the primary
star, appears $5.3$ kyr after the start of the simulation. It forms in
a shocked filament, which continues to accrete mass and by $6$ kyr is
beginning to form a flattened protostellar disk. Row (2), shows the
state of the simulation at this point.

\begin{figure*}
\plotone{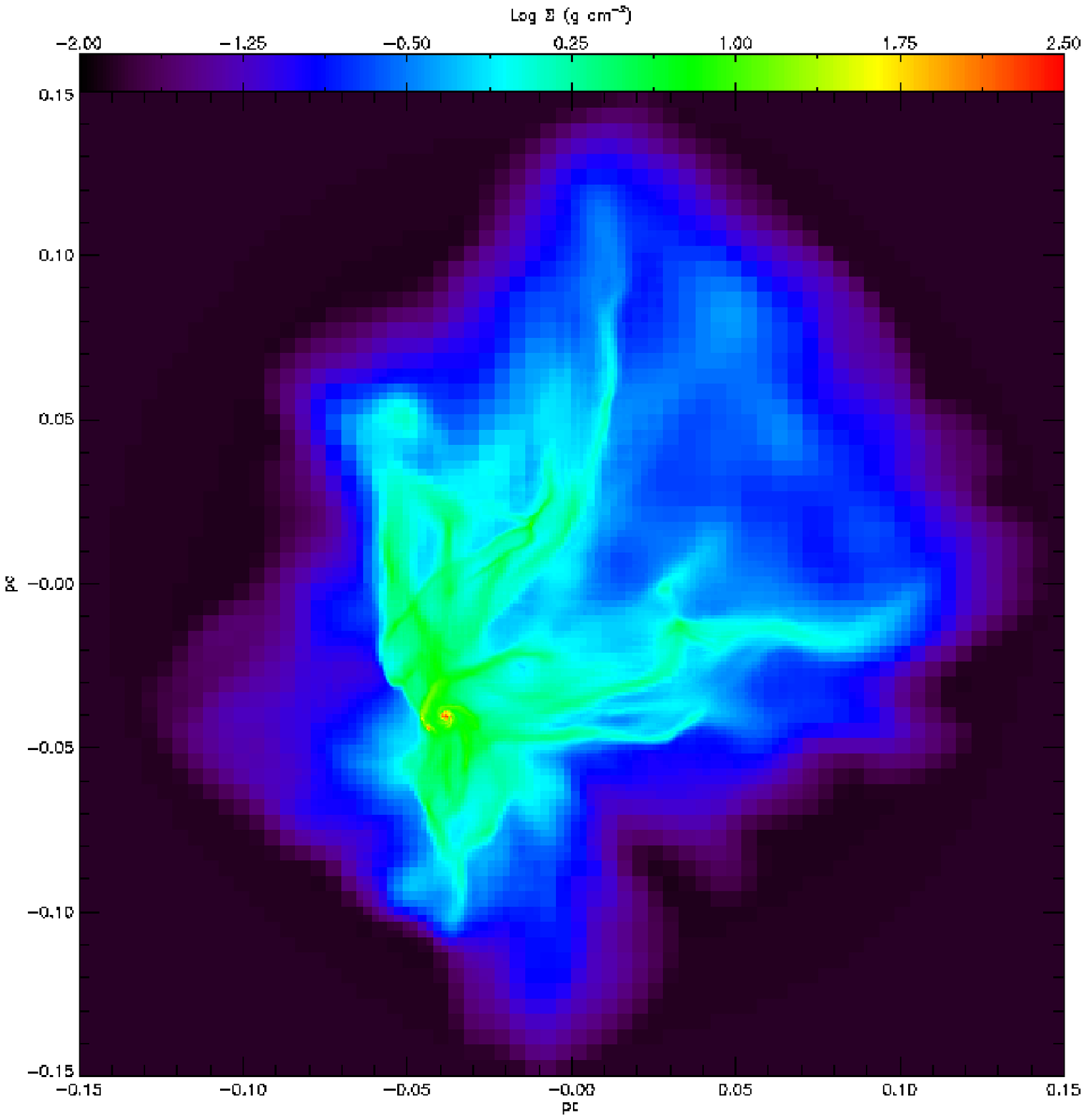}
\caption{
\label{100Afull}
Column density at 20 kyr in run 100A. We do not show the positions of
stars.\\
}
\end{figure*}

As the evolution continues, several more dense condensations appear,
but most of these are unable to collapse and form a protostar before
reaching the primary star and being sheared apart in the protostellar
disk.
At $12.2$ kyr a second protostar forms, but it falls into
the primary star and merges with it at $12.7$ kyr, before it has
accreted $0.1$ $\msun$ of gas. At the time of merging the primary star
is already $2.1$ $\msun$, so the mass gained in the merger is
negligible. Row (3) shows the state of the simulation at $12.5$ kyr,
about halfway between when the second star appears and when it merges
with the primary. We should at this point mention a caveat regarding
resolution. Due to our limited resolution, we are unable to resolve
binaries in orbits closer than 8 cells, or 60 AU. This means that it
is likely that at least some of the mergers identified by our code are
really formations of tight binaries. We discuss the implications of
this in greater detail in our discussion of numerical resolution
issues in \S~\ref{resolution}. Here, we simply mention that whether
the true outcome is a merger of a tight binary probably makes very
little difference to the overall evolution, since the smaller star
carries negligible mass compared to the primary.

Only after $14.4$ kyr does one of the condensations collapse to form a
second protostar that is not immediately accreted, as shown in row
(4). At this point the primary star is $3.2$ $\msun$, and has a
well-defined massive disk. The condenstation from which the new
protostar forms is already visible in row (3). It is able to
collapse and form a protostar, unlike several others, because it is
fairly distant from the central object. This reduces the amount of
radiative heating to which it is subjected, a topic we discuss in more
detail in \S~\ref{radheating}.

The next significant change in the system occurs when one of the arms
of the disk becomes unstable and fragments to form a third protostar
at $17.4$ kyr. We show the configuration just after this in row
(5). At this point the central star
mass is $4.3$ $\msun$. The disk mass cannot be defined precisely,
because the disk does not have a clearly defined edge. However, we can
get an approximate mass by defining the disk as all the cells within
1000 AU of the primary star denser than $10^{-15}$ g cm$^{-3}$. This
gives a mass range of $3.1$ $\msun$. We discuss our definition for the
disk edge, and disk fragmentation in general, in
\S~\ref{disks}. The fragment is very small compared to the
central protostar, and remains so as the simulation continues to
evolve.

The configuration after 20 kyr of evolution, shown in row (6), is
substantially similar. Two more small disk fragments form, but they
both collide with the primary star almost immediately after formation,
when their masses are $<0.1$ $\msun$. This has a negligible effect on
the mass of the primary. At the end of 20 kyr, the primary star is $5.4$
$\msun$, the second star is $0.34$ $\msun$, and the third star, which
formed in the disk of the first, is $0.20$ $\msun$. The disk itself is
$3.4$ $\msun$ in mass. Thus, the system is well on
its way to forming a massive star, and thus far the vast majority of
the collapsed mass has concentrated into a single object. We show a
larger plot of the full core at this point in Figure \ref{100Afull}.

\begin{figure}
\plotone{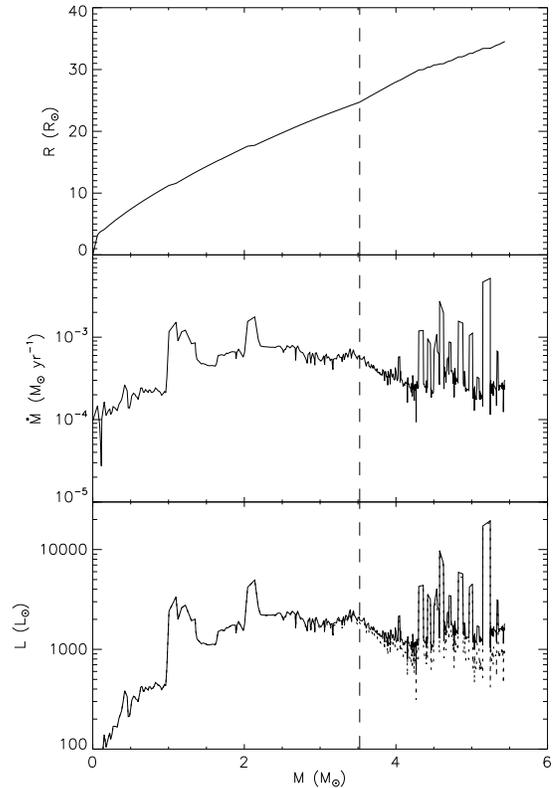}
\caption{
\label{starhist}
Radius, accretion rate, and luminosity of the primary star versus mass
in run 100A. The dashed vertical line indicates the mass at which the
star begins burning deuterium. In the luminosity plot, the solid line
is total luminosity and the dotted line is the luminosity due to
accretion. The two are identical before deuterium burning begins.
\\
}
\end{figure}

We show the evolution of the primary star's radius, mass accretion
rate, and luminosity in Figure \ref{starhist}. The model of
MT03 gives an accretion rate onto the star-disk system of
$1.2\times 10^{-4}(m_{*d}/1\, M_\odot)^{1/2}$ $\msun$ yr$^{-1}$, where
$m_{*d}$ is the mass of the star plus the disk. If we assume that the
accretion rate onto the star is a fraction $m_*/m_{*d}$ of this, we
infer an accretion rate of about $2\times 10^{-4}$ $\msun$ yr$^{-1}$
onto the star, which is comparable to the simulation result at the end
of the calculation; at earlier times, the simulation gives a higher
accretion rate. There are at least three reasons for this. First, we
assume that the density in the central regions was initially
constant, whereas MT03 assume that the power-law
increase in density continues all the way to the origin; as a result,
accretion is delayed for 6 kyr in the simulation, whereas it begins
immediately in the analytic model, thereby spreading the accretion
over a longer time. Second, MT03 assume that the
turbulence is undamped, whereas it is damped in somewhat less than a
crossing time in the simulation, which increases the accretion
rate. Third, our primary star forms out of a single large, shocked
filament, and accretion onto it is dominated by gas from this
filament. The shock raises the density and thereby enhances the
accretion rate relative to what the MT03 model
predicts.

\subsubsection{Run 100B}

\begin{figure*}
\plotone{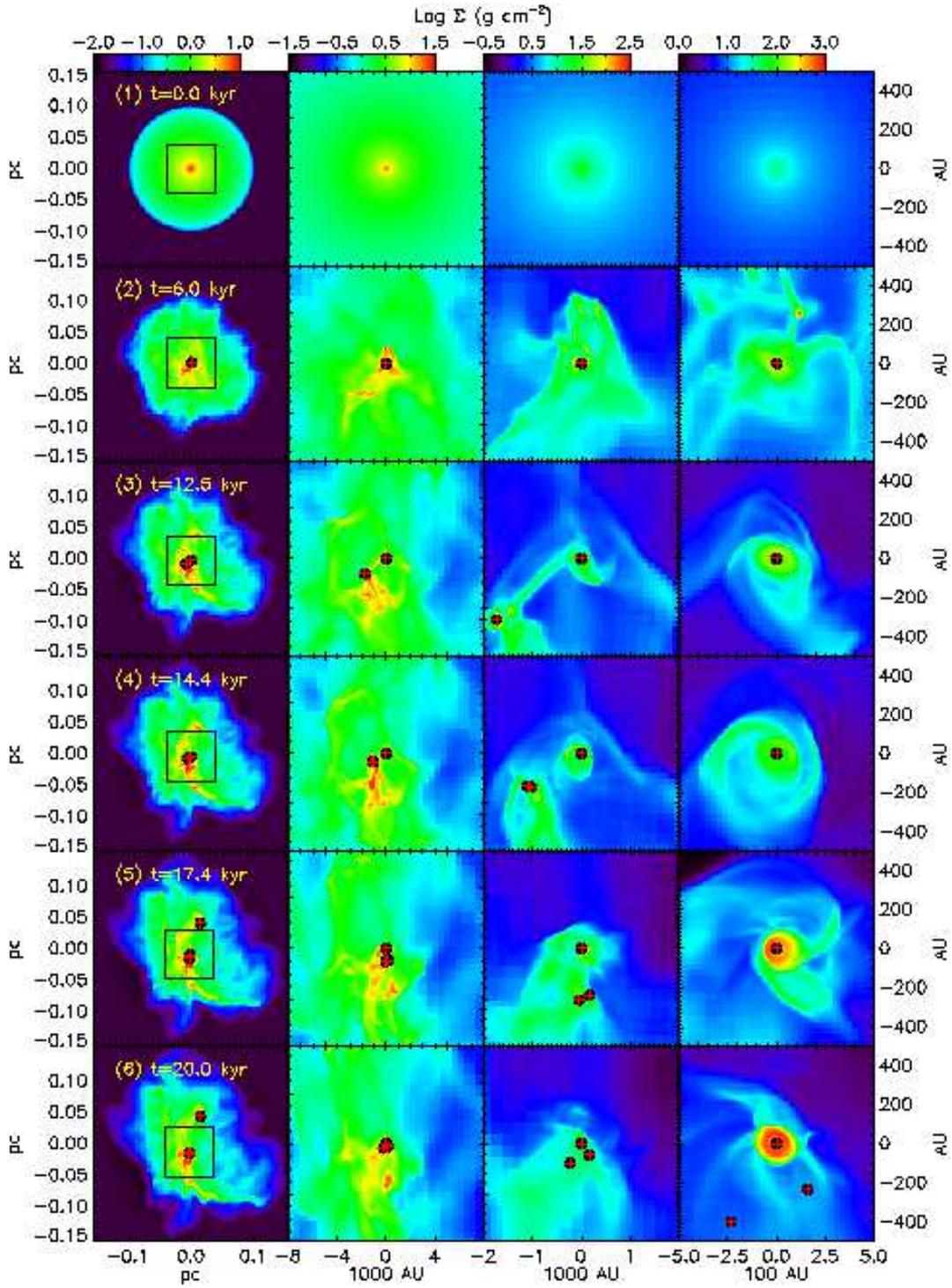}
\caption{
\label{100Bevol}
Column density as a function of time in run 100B. See Figure
\ref{100Aevol} for a detailed description.\\
}
\end{figure*}

Run 100B uses the same parameters as run 100A, but a different random
realization of the initial turbulent velocity field. The evolution in
detail is of course different than for run 100A, but qualitatively it
is quite similar. Figure \ref{100Bevol} shows a time sequence of
column densities at the same times as in run 100A. On the largest
scales, the as in run 100A, at late times the core is dominated by
large-scale filaments. There is a primary star at the center of this
filament network, and on smaller scales it is surrounded by a
disk. the disk is somewhat denser and less extended than in run 100A,
but differs in size by less than a factor of 2.

In run 100B, a primary star forms at 5.4 kyr, and a second object
forms at 7.8 kyr. However, the two merge at 9 kyr, when the primary is
$2.0$ $\msun$ and the secondary is only $0.17$ $\msun$. By 12.5 kyr,
as Figure \ref{100Bevol} shows, the primary is surrounded by a
well-defined flattened disk. Accretion onto the primary slows down
after that point, and the disk around it remains fairly stable,
until the primary collides with a second star at 16.1 kyr. This
increases the primary's mass from 4.2 $\msun$ to $5.1$ $\msun$, and
the accretion rate increases thereafter, since the collision brings in
an amount of gas considerably larger than the amount of mass the
primary gains by the collision itself. This also reduces the angular
momentum in the disk around the primary, allowing more gas to accrete
and the disk to become more compact. As a result of these two effects,
the primary grows from $5.1$ $\msun$ to $8.9$ $\msun$ in the 4 kyr
after the collision, gaining roughly three times as much mass from gas
accretion as from the collision.

The disk around the primary looks similar to that in run 100A at times
before the collision at 16.1 kyr, including the beginnings of spiral
structure. However, after the collision the disk is denser and more
compact, and lower in mass smaller relative to the star. At 20 kyr,
the disk mass is $2.4$ $\msun$, roughly a quarter of the mass of the
star. Due to its smaller mass and radius, the disk appears to be more
stable than the disk in run 100A, and there is no evidence for disk
fragmentation. Since this is a result of the collision, an event that
shows no signs of being repeated, it seems likely that further
evolution will cause the disk to become larger again, and return it to
a state of instability similar to that of the disk in run 100A.

\subsubsection{Run 200A}

\begin{figure*}
\plotone{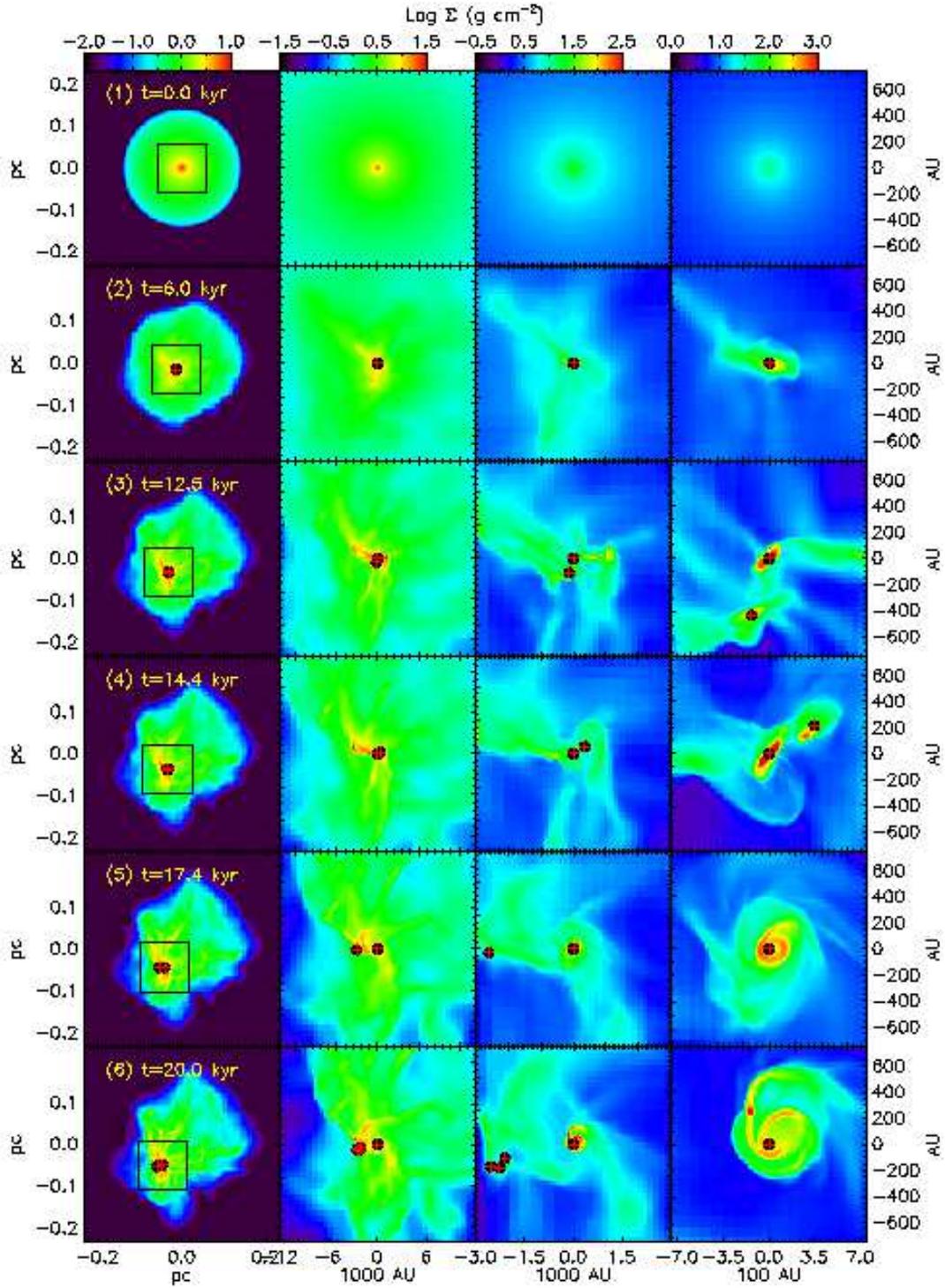}
\caption{
\label{200Aevol}
Column density as a function of time in run 200A. See Figure
\ref{100Aevol} for a detailed description. Note that, to accomodate
the somewhat larger size of the core, the areas shown in each panel
are a factor of 1.5 larger in linear dimension than the analogous
panels in \ref{100Aevol}.\\
}
\end{figure*}

In run 200A we use the same turbulent velocity field as in run 100A,
but imposed on a $200$ $\msun$ core with the same initial column density
as the $100$ $\msun$ core in run 100A.
Figure \ref{200Aevol} shows the time sequence of column densities in
run 200A at the same times as in run 100A. The overall appearance of
the core is very similar to in run 100A, as are the positions and
times of formation of the protostars. This is not surprising given the
identical initial velocity fields. A primary star forms at 5.3 kyr,
and a second at 11.5 kyr, but this second one merges with the primary
at 15.1 kyr. At 16.1 kyr, a second fragment forms
at a larger distance from the primary, and it survives. In the final
time step, another pair of small stars form near it. From $\sim 10$
kyr of evolution onward, the primary has a disk extending several
hundred AU out. 

At the end of the run, the primary star has reached $8.6$ $\msun$, and
the accretion disk around it is roughly $6$ $\msun$. As in run 100A,
there is obvious spiral structure in the disk. Unlike in run 100A,
there is no disk fragmentation, although in the final time slice one
sees a dense condenstation that is analogous to the one that formed a
protostar out of the disk in run 100A, and looks likely to produce a
fragment. The accretion rate onto the star is roughly $5\times
10^{-4}$ $\msun$ yr$^{-1}$. This agrees well with the MT03 model,
which for the initial conditions in run 200A predicts an accretion
rate of $6.5\times 10^{-4}$ $\msun$ yr$^{-1}$ onto a $14.6$ $\msun$
star-disk system.

\subsubsection{Run 100ISO}

\begin{figure*}
\plotone{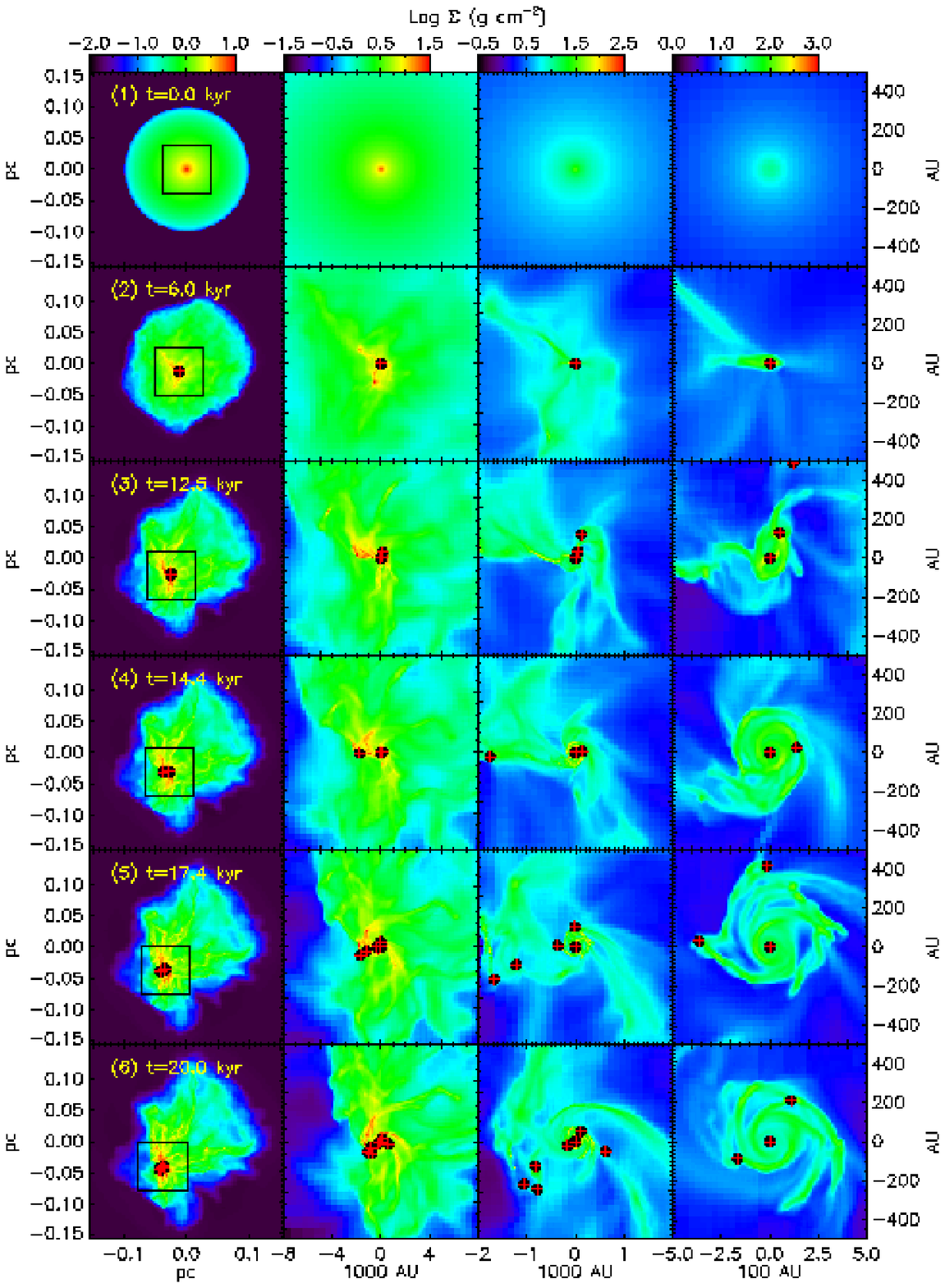}
\caption{
\label{100ISOevol}
Column density as a function of time in run 100ISO. See Figure
\ref{100Aevol} for a detailed description.\\
}
\end{figure*}

We show a time sequence for the evolution of the isothermal run at the
same times as for run 100A in Figure \ref{100ISOevol}. The overall
morphology of the gas is fairly similar on large scales, which is not
surprising given the identical initial conditions. A first
condensation forms, and 2 kyr after the start of the run a primary
object forms; it remains the most massive star throughout the
run. However, on smaller scales runs 100A and 100ISO show very
significant differences. Due to its lack of thermal support compared
to run 100A, the gas in the isothermal run is much more
filamentary. Disks are flatter, filaments have smaller radii, and
shock structures are thinner. This causes the evolution to proceed
quite differently, so that by 20 kyr it is fairly difficult to line up
the features from the two runs except at the grossest level of the
location of protostellar disks and major filaments. There are many
more fragmentation sites, and objects are generally much clumpier than
in the radiative run. The number of protostars is considerably
larger.

\subsection{Statistics of Fragment Formation}
\label{fragstats}

\begin{figure}
\plotone{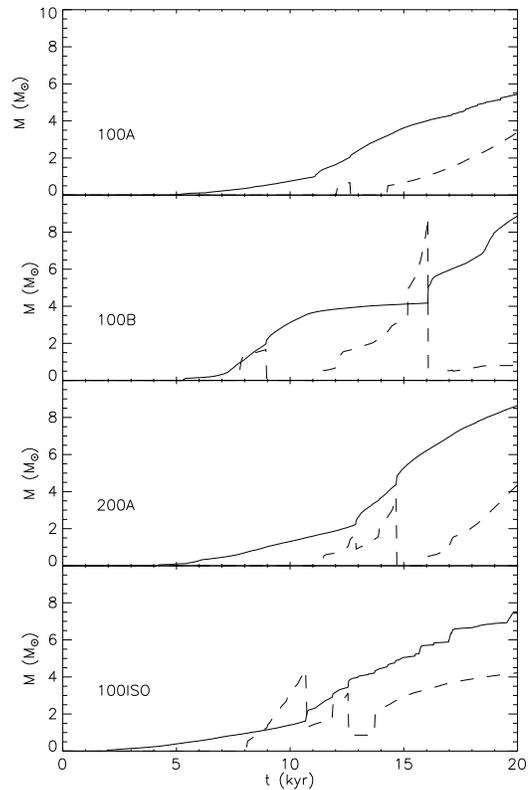}
\caption{\label{starmass}
Mass of most massive star (\textit{solid line}) and ten times the mass
of the second most massive star (\textit{dashed line}) as a function
of time in all runs. Sudden increases in mass correspond to points
where a smaller star merges with a bigger one. Sudden decreases in
mass correspond to the points where the title ``second most massive''
star suddenly changes from one star to another because the previous
second most massive star has merged with the most massive.\\
}
\end{figure}

We summarize the statistics of the stars that form in our simulations
in Table \ref{starstatistics}. In addition to describing the total
number and mass of stars present at the end of the run, we give the
total number of stars formed, the total number of significant stellar
mergers, defined as those that alter the mass of the more massive
merger companion by at least 5\%, the final mass of the primary and
of all the other stars combined, and the fraction of the primary's
mass acquired by mergers rather than accretion, defined as the total
mass of all stars that merge with the primary immediately before the
merger, divided by the primary's final mass. We show the evolution
of the mass of the most massive and second most massive stars in all
the runs in Figure \ref{starmass}.

The statistics and plot make clear that in all the radiative runs
the primary gains mass almost exclusively by accretion. The number of
fragments formed is small, and the number surviving at the end of the
run even smaller. Moreover, these fragments always contain an extremely
small fraction of the total collapsed mass. In none of the radiative
runs does the primary star gain more than $\sim 10\%$ percent of its mass by
collisions. There are kinks in the mass versus time curves shown in
Figure \ref{starmass}, indicating sharp rises in the accretion rate,
but most of these are due to the primary encountering and accreting
dense gas condensations that had not formed stars, not due to
mergers. In summary, fragmentation appears to be very weak in massive
protostellar cores once we take into account radiative feedback, and
stars appear to gain mass by accretion rather than by collisions.

The fragmentation history is very different in run 100ISO.
At 20 kyr, there are 7 protostars in run 100ISO, and opposed to
3 in run 100A. There are several more that we do not list because they
are just below the $0.05$ $\msun$ cutoff. In the isothermal run,
these are certainly real, since there are no potential problems with
an iterative radiation solver. Moreover, a factor of 4 more protostars
form over the course of run 100ISO than in run 100A, and the fraction
of its mass that the primary gains by merging rather than accretion is
nearly an order of magnitude larger. As a result the plot of
primary mass versus time shown in Figure \ref{starmass} is much
spikier. Some of the additional stars form
out of the disk around the primary, while others form in separate
condensations. In several cases we can identify analogous
condensations in the isothermal and radiative runs. In the radiative
run these are too hot to collapse, and instead they reach the primary
and are accreted, but in the isothermal run they collapse and form
protostars.

The fragmentation we see in run 100ISO is mostly consistent with
previous work on fragmentation in isothermal simulations, which
generally find a great deal of fragment formation. \citet{dobbs05}
find that a simulation of a $30$ $\msun$ turbulent core forms
$\sim 30$ fragments over an evolution time slightly longer than
ours. They do not find any massive stars forming, but that is likely
because our effective resolution is considerably lower than theirs
(see \S~\ref{resolution}), so they have fewer mergers and more
fragment formation around their central object. Simulations with
adaptive SPH codes find that, for isothermal equations of state, the
amount of fragment formation and the typical fragment mass are both
highly resolution-dependent \citep{martel06}. Runs with radiative
transfer are unlikely to suffer from this problem, because radiative
heating shuts off fragmentation on small scales. (It is worth noting
that we could go to higher resolution in the isothermal run, since it
is computationally cheaper by a factor of $\sim 6-7$ than runs with
radiative transfer. However, to make the comparison as fair as
possible, we use the same resolution in the radiative and
non-radiative runs.)

\subsection{Radiative Heating and Fragmentation}
\label{radheating}

Clearly there is a significant difference
between the radiative transfer runs, none of which show much
fragmentation and for which the morphology is relatively smooth, and
the isothermal run, in which numerous fragments form out of a strongly
filamentary morphology and a significant fraction of the primary's
mass is acquired through mergers. This
suggests that the effective equation of state, including radiative
heating, is playing an important role in determining how fragmentation
occurs. Examining the distribution of temperature and Jeans mass,
\begin{equation}
M_J = \rho \lambda_J^3 = \left(\frac{\pi k_B T}{G \mu}\right)^{3/2}
\rho^{-1/2},
\end{equation}
in our simulations supports this hypothesis. For reference, in the
initial state the Jeans mass at the core edge is $3.4$ $\msun$, at the
mean density it is $2.4$ $\msun$, and at the central density it is
$0.03$ $\msun$. Note that this Jeans mass is 4.71 times the
Bonnor-Ebert mass, so centrally-condensed objects with masses
considerably smaller than $M_J$ can still be unstable.

\begin{figure}
\plotone{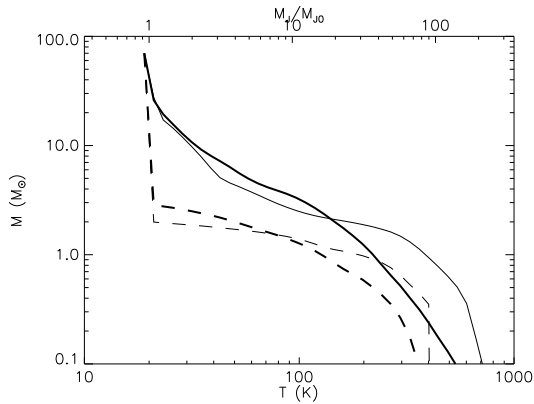}
\caption{
\label{100Atdist}
Mass of gas above temperature $T$ as a function of
$T$ in run 100A. The curves do not reach $100$ $\msun$ because this
analysis only includes gas with a density more than twice the initial
density at the edge of the core, to ensure that there is no confusion
with the ambient medium. We show the state of the run at $12.5$ kyr, just
after the second fragment forms (\textit{thin solid line}), and at the
end of the run, $20$ kyr (\textit{thick solid line}). We also show the
distributions at those times computed using temperatures derived from
a barotropic equation of state rather than the true temperatures in
our run (\textit{thin and thick dashed lines}). The top axis shows the
ratio of Jeans mass $M_J$ at temperature $T$ to Jeans mass $M_{J0}$ in
gas of the same density at the initial temperature $T_0=20 K$. For gas
at the mean density of the initial core, $M_{J0}=2.4$ $\msun$.
\\
}
\end{figure}

\begin{figure*}
\plotone{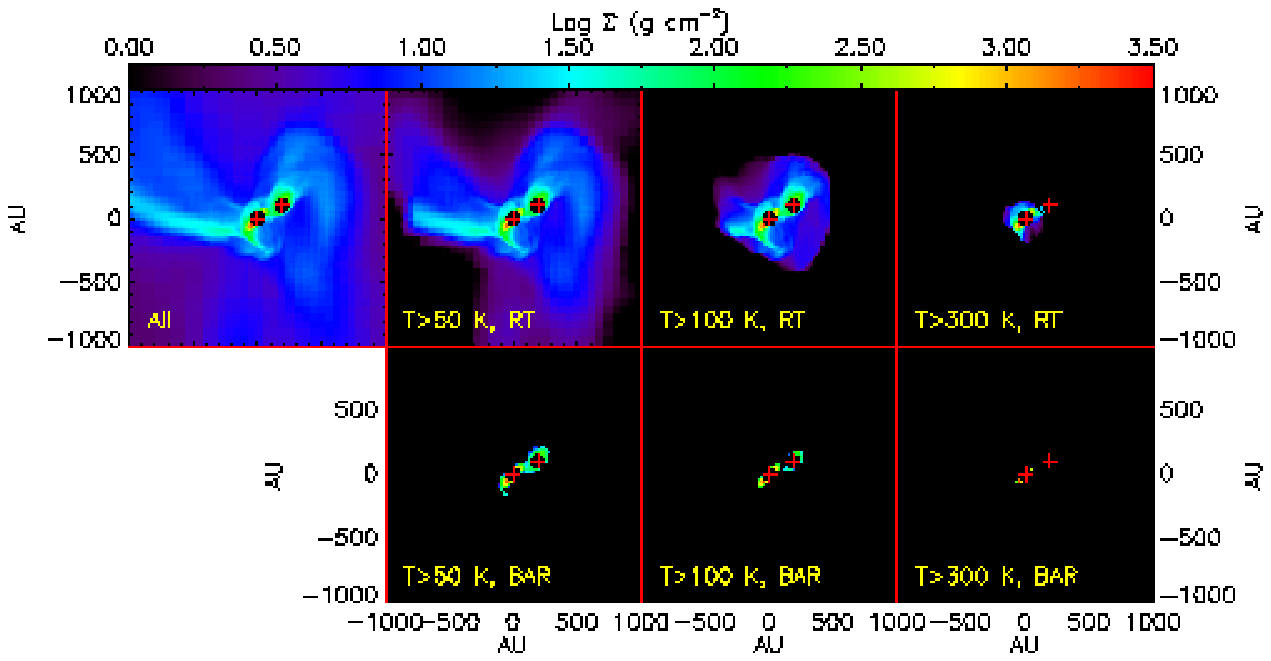}
\caption{
\label{100Atemp}
Column density in run 100A of gas above the temperature indicated in
each panel. The top left panel shows all the gas ($T>0$), and the top
row shows gas above temperatures of 50 K, 100 K, and 300 K. The bottom
row shows the column density above those temperatures, but using
temperatures computed from a barotropic equation of state rather than
the actual temperature in the run. Stars are indicated by white plus
signs. The time shown is the same one shown in Figure \ref{100Aevol},
row (3): $12.5$ kyr, shortly after the time the second protostar
forms.
\\
}
\end{figure*}

In Figure \ref{100Atdist} we plot for run 100A the mass $M(>T)$ of gas
with temperature greater than $T$, at a time shortly after the first
fragment other than the primary star forms ($12.5$ kyr) and at the final
time in the run ($20$ kyr), and in Figure \ref{100Atemp} we show the
spatial distribution of the gas as a function of temperature at 12.5
kyr. Clearly by the
time the second star forms, radiation from the primary has heated a
significant fraction of the core to well above its initial
temperature. The temperature is above 50 K in $4.4$ $\msun$ of gas,
including almost all the gas within $1000$ AU of the primary object,
which is where much of the fragmentation takes place in our isothermal
run and in other isothermal simulations in the literature
\citep[e.g.][]{bate03}. By 20 kyr, the mass heated to more than 50 K
is $6.0$ $\msun$, extending more than 2000 AU away from the primary star.

\begin{deluxetable*}{ccccccc}
\tablecaption{\label{starstatistics}Statistics of Stars Formed}
\tablehead{
Run & $N_{20}$ & $N_{\rm formed}$ & $N_{\rm merge}$ & $M_{1}$ $(\msun)$ & 
$M_{\rm other}$ $(\msun)$ & $f_{\rm merge}$
}
\startdata
100A & 3 & 6 & 0 & 5.4 & 0.54 & 0.04 \\
100B & 4 & 7 & 3 & 8.9 & 0.31 & 0.12\\
200A & 4 & 6 & 2 & 8.6 & 0.54 & 0.06 \\
100ISO & 7 & 23 & 6 & 7.4 & 1.5 & 0.31 \\
\enddata
\tablecomments{Col. (2): Number of stars present at $20$
kyr. Col. (3): Total number of stars formed over the $20$ kyr
evolution, including those that have merged. Col. (4): Number of
significant merger events. Col. (5): Mass of primary
star to 20 kyr. Col. (6): Total mass of all stars but the primary at 20
kyr. Col (7): Fraction of primary's mass acquired by mergers.
}
\end{deluxetable*}

\begin{figure}
\plotone{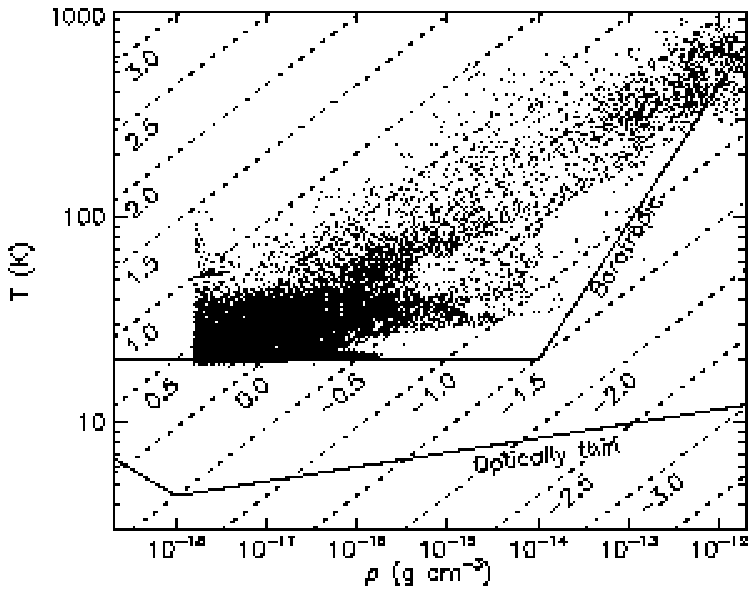}
\caption{
\label{100Arhot}
Scatter plot showing temperature versus density for a sample 50,000
cells in run 100A at time $12.5$ kyr. Cells are selected with a
probability proportional to their mass, so the density of points is a
true representation of the mass distribution, with the exception that
we exclude cells with densities below twice the initial cloud edge
density to ensure that we exclude the ambient medium. 
The diagonal dotted lines are curves of constant Jeans
mass. The number next to each line indicates the
value of $\log (M_J/\msun)$ to which it corresponds. The solid lines are
the curves of $\rho$ versus $T$ for the barotropic equation of state of
\citet{dobbs05} and for the optically thin heating and cooling model of
\citet{larson05}.
\\
}
\end{figure}

To study how this is heating is likely to affect fragmentation, in
Figure \ref{100Arhot} we show a scatter plot of density versus
temperature for the mass in run 100A at 12.5 kyr. From the plot, it is
clear that almost all of the very dense gas, where fragmentation might
take place, is heated to hundreds of K. However, this heating begins
at relatively low densities, so almost all the gas denser than 10
times the initial density has been heated at least somewhat. The
$\rho-T$ distribution is roughly bounded by $T\propto \rho^{\gamma-1}$
with $\gamma=1.2-1.3$ over the full range of the density
distribution. The exact shape changes at other times, but the general
feature that the temperature rises continuously with density, with no
large isothermal density range, persists at all times after the
primary object forms.

Since the rise is slower than $T\propto
\rho^{1/3}$, the Jeans mass does decline as one moves to higher
density material, and so except at the highest densities and
temeperatures there is generally more than a thermal Jeans mass of
material above any given density. At the mean density and temperature of
the core, there are still many thermal Jeans masses. However, the
continuous rise of
$T$ with $\rho$ in our core still provides an explanation why we see
so little fragmentation. Simulations
\citep{li03,jappsen05,bonnell06d} and analytic work \citep{larson05}
suggest that how fragmentation in a turbulent medium proceeds depends
critically on the value of $\gamma$, with fragmentation proceeding to
arbitrarily small masses as long as $\gamma < 1$, and ceasing for
$\gamma > 1$. The physical argument behind this result is that, in a
turbulent medium, the densest structures formed by the turbulence are
generally filaments. Gravitationally unstable filaments are able to
collapse axisymmetrically toward their centers when $\gamma < 1$, but
are unable to collapse axisymmetrically for $\gamma>1$. If the rate of
radiative heating and cooling has a density dependence such that
$\gamma<1$ at low densities and $\gamma>1$ at high densities, then
fragments will form at the density corresponding to the transition
between these two, because at this density contraction of filaments
along their axes will stall, and the filament will break up into
``beads'' instead.

Figure \ref{100Arhot} shows that in a massive core with
an accreting protostar in its center, $\gamma>1$ effectively over the
entire core. There is a region of points with $\gamma\approx 1$, in
the form of the line of points at $T\approx 30$ K at densities from
$10^{-16}-10^{-14}$ g cm$^{-3}$, and indeed these points do represent
the gas from which the next fragment forms, at 14.4 kyr of
evolution. They are relatively cool because they are $\sim
3000$ AU from the primary star, and are sufficiently dense to be
self-shielding against its radiation. However, these points are the
exception, and overall such self-shielding distant structures form
only rarely. This is likely why we see so little fragmentation despite
the fact that our simulation contains many tens of thermal
Jeans masses. Filaments do form, but they are unable to even begin
contracting because radiative heating keeps them at $\gamma >
1$. Rather than contracting, stalling, and breaking up into beads,
they never begin contracting in the first place, and instead their
mass drains onto the primary star or its disk.

It is important to note that the energy source in our simulation
responsible for raising the temperature is almost entirely accretion
onto the primary star. At $12.5$ kyr the primary star has not yet
started burning deuterium and is only 2 $\msun$, so the luminosity all
comes from accretion. At $20$ kyr deuterium has ignited but is only
generating $\sim 50\%$ of the total luminosity. Thus, the heating does
not depend on nuclear burning in the primary star. Accretion
luminosity by itself is sufficient to greatly reduce
fragmentation. However, as the luminosity rises due to nuclear
burning, the effect should become even more significant.

For comparison, in Figures \ref{100Atdist} and \ref{100Atemp} we also
show results using the density distribution in our simulation, but
using temperatures computed from a barotropic equation of state rather
than the real temperatures in the simulation. Simulations have used a
variety of barotropic equations of state. We compare to one from
\citep{dobbs05}, which generally produces higher temperatures than
those used elsewhere \citep[e.g. by][]{bate03, li03, jappsen05},
\begin{equation}
\label{bareos}
T(\rho) = T_0 \cdot \left\{\begin{array}{cc}
1, & \rho < \rho_0 \\
(\rho/\rho_0)^{2/3}, & \rho_0 \le \rho \le \rho_1 \\
(\rho_1/\rho_0)^{2/3}, & \rho > \rho_1
\end{array}\right.,
\end{equation}
with $T_0=20$ K, $\rho_0=10^{-14}$ g cm$^{-3}$ and $\rho_1=10^{-12}$ g
cm$^{-3}$. As is clear from the Figures, the barotropic equation of
state severely underestimates both the temperature of the gas
and the spatial extent of the heated region.

In Figure \ref{100Arhot} we show both the $\rho-T$ curve resulting
from the \citet{dobbs05} barotropic equation of state, and also the
curve of the optically thin heating and cooling model of
\citet{larson05},
\begin{equation}
T(\rho) = T_0  \cdot 
\left\{\begin{array}{cc}
(\rho/\rho_0)^{\gamma_0-1}, & \rho < \rho_0 \\
(\rho/\rho_0)^{\gamma_1-1}, & \rho \ge \rho_0 \\
\end{array}\right.,
\end{equation}
with $T_0 = 4.4$ K, $\rho_0 = 10^{-18}$ g cm$^{-3}$, $\gamma_0=0.73$,
and $\gamma_1=1.07$. Clearly this equation of state underestimates the
temperature even more severely than the \citeauthor{dobbs05}
barotropic equation of state. For either the \citeauthor{dobbs05} or
\citeauthor{larson05} equations of state, the Jeans mass in our
simulation is larger than the Jeans mass they would predict, often by
orders of magnitude. Thus, regardless of the details of how
fragmentation occurs, we expect that simulations that adopt either of
these proposed equations of state will overpredict the number of
fragments. Moreover, both the barotropic and optically thin
equations of state produce a range of density with $\gamma\le 1$ where
the thermodynamics favor fragmentation, while our simulation shows
that radiation feedback largely prevents this type of thermodynamic
behavior.

It is not surprising that the barotropic and optically thin cooling
equations of state fare so poorly. As first pointed out by
\citet{krumholz06b}, in a collapsing protostellar core, before nuclear
burning starts the largest energy source either internal or external
to the core is the gravitational potential energy released in the
final plunge of gas onto the stellar surface. As a result, unless one
explicitly includes the energy released by accretion onto the
protostellar surface, and the radiative transfer of this energy to the
rest of the core, one is ignoring the dominant source of energy in the
problem. This is exactly what the barotropic and optically thin
approximations do. Our results indicate that this is likely to result
in qualitatively incorrect results for fragmentation in massive
cores.

Nor can the problem be fixed simply by using better approximate
equations of state. As the Figures show, the temperature distribution
is a function of both time and space, and can change in unexpected
ways. For example, despite the fact that the luminosity is comparable
at 20 kyr to that at 12.5 kyr, and there is more gas
heated to moderate temperatures $\sim 50-100$ K, there is actually
less mass at temperatures $\gg 100$ K. This is largely because an
optically thick disk has formed which is shielding much of the dense gas
from protostellar radiation. No equation of state that gives the
temperature simply as a function of the density or other local gas
properties will reproduce effects like this.

\subsection{Disk Properties}
\label{disks}

\begin{figure*}
\plotone{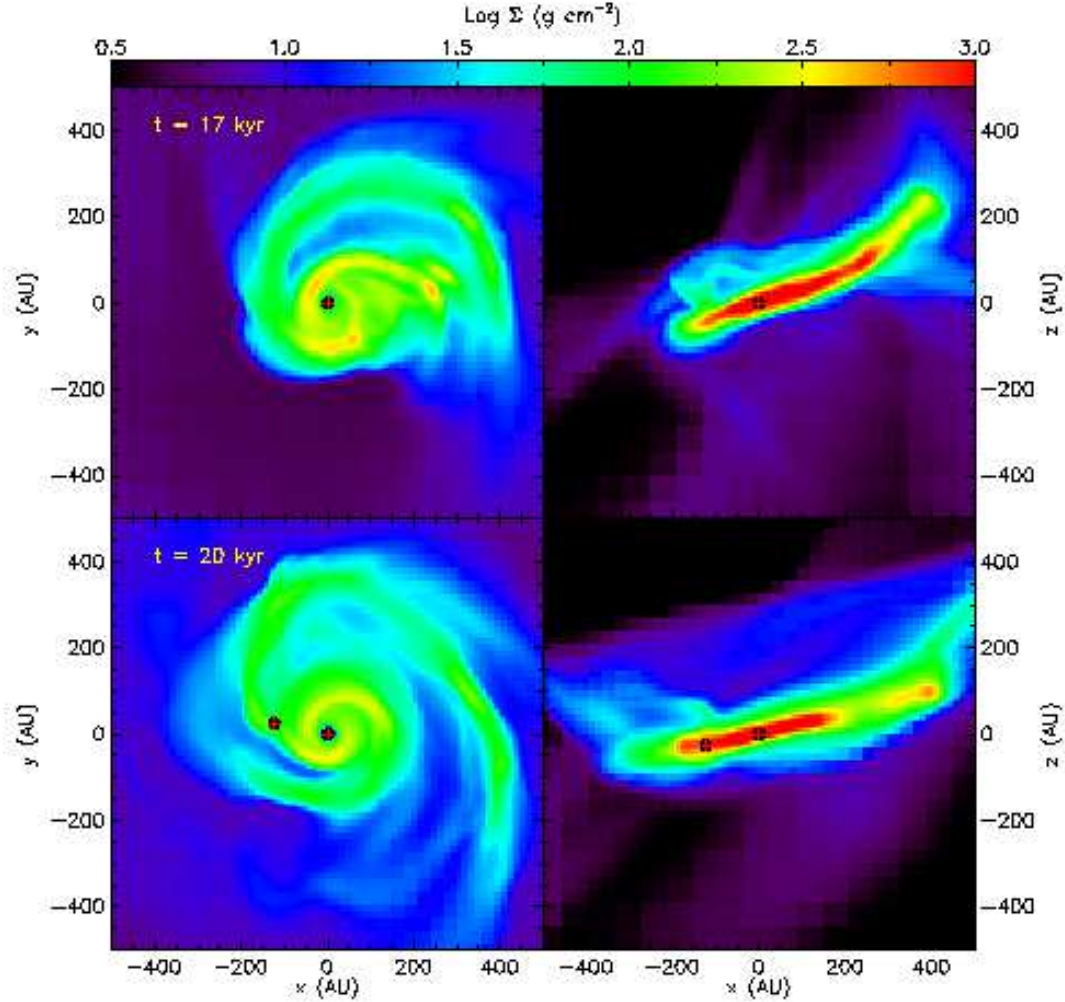}
\caption{
\label{100Adiskimages}
Column density of the disk in run 100A at 17 kyr (\textit{top row})
and 20 kyr (\textit{bottom row}), in two orthogonal projections. Stars
are indicated by white plus signs.
\\
}
\end{figure*}

Here we analyze the properties of the disk around the primary star
that forms in run 100A, to better understand both angular momentum
transport and disk fragmentation. To examine the properties of the
disk, we must first isolate it from the background flow. To do so, we
choose a density threshold of $10^{-15}$ g cm$^{-3}$ to separate disk
material from the ambient gas. This threshold agrees reasonably well
with what one identifies by eye, and values
up to a factor of $\sim 10$ different from this do not produce
qualitatively different results. We also focus on gas within 1000 AU
of the primary, to ensure that the gas we are examining is in orbit
around it rather than around other protostars. After using these
criteria to remove extraneous gas, we compute the total
angular momentum about the primary of the remaining gas.
Since our disk is not aligned with the computational grid, to
analyze it we ``deproject'' by computing the column density
$\Sigma$, mass-weighted sound speed $c_s$, and mass-weighted angular
velocity $\Omega$ projected onto the plane orthogonal to the angular
momentum vector. We do this at 17 kyr, just before the first disk
fragment forms, and at 20 kyr, the end of our run. The deprojected
maps of disk properties at these times form the basis of our
analysis. We show the disk at these two times, before deprojection, in
Figure \ref{100Adiskimages}.

\subsubsection{Angular Momentum Transport}

First we wish to determine the effective viscosity $\alpha$ for our
disks. In a Keplerian disk this is related to the kinematic viscosity
$\nu$ by $\nu=2 \alpha c_s^2/(3\Omega)$ \citep{shakura73}. The inward
radial drift velocity of the material in the disk is $v_R\approx \nu/R$,
and the accretion rate onto the central star is $\dot{M}=2\pi R \Sigma
v_R$, so $\alpha\approx \dot{M}\Omega/(3\pi\Sigma c_s^2)$. We
therefore estimate $\alpha$ by computing the mass-weighted average
\begin{equation}
\left\langle \alpha\right\rangle 
\approx 
\frac{
\int_{-\pi}^{\pi} d\phi \int_{0}^{\infty} dr \,
\Sigma \frac{\dot{M} \Omega}{3\pi \Sigma c_s^2}
}{
\int_{-\pi}^{\pi} d\phi \int_{0}^{\infty} dr \,
\Sigma
}.
\end{equation}
For the properties of our disk at either 17 kyr or 20 kyr, and taking
$\dot{M}$ to be the mean accretion rate between these times, $4\times
10^{-4}$ $\msun$ yr$^{-1}$, we find an effective $\alpha \approx
1.0-1.6$. This is quite rapid angular momentum transport, and is
significantly larger than what one expects due to purely local
transport phenomena \citep[e.g.][]{gammie01}. Accretion is obviously
highly time-dependent and unsteady, and the disk never settles into a
steady state, so the rate of angular momentum transport at any instant
may be very different from the average.

\begin{figure}
\plotone{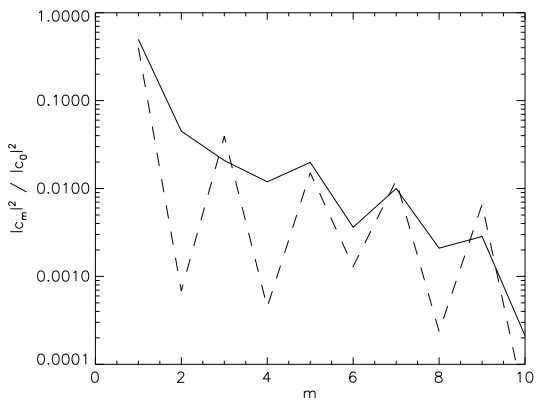}
\caption{
\label{100Aspiral}
Normalized power $|c_m|^2/|c_0|^2$ in azimuthal mode $m$ in the disk
around the primary star, at 17 kyr (\textit{solid line}) and 20 kyr
(\textit{dashed line}).
\\
}
\end{figure}

To understand the angular momentum transport mechanism, we
analyze the spiral pattern in the disk by computing Fourier
coefficients of the density distribution in azimuth around the primary
star. Defining $r$ as the distance from the primary star in our
projection and $\phi$ as an angular coordinate in the projection
plane, we compute
\begin{equation}
c_m = \frac{1}{2\pi} \int_{-\pi}^{\pi} d\phi \int_{0}^{\infty} dr \,
e^{i m \phi} r \Sigma(r,\phi).
\end{equation}
We plot the normalized power $|c_m|^2/|c_0|^2$ for $m=1-10$ in Figure
\ref{100Aspiral}. As the figure shows, at both 17 kyr and 20 kyr the
vast majority of the power is in the $m=1$ spiral mode, with smaller
amounts of power in other odd modes and very little power in even
modes. This suggests that angular momentum transport and spiral arm
formation in our disk is primarily due to the SLING instability
\citep{adams89, shu90}. For disks as massive as ours, $M_d\sim 0.5
M_*$, this mechanism enables accretion on a disk dynamical time scale
rather than a viscous time scale, consistent with our value of
$\alpha\sim 1$. Angular momentum transport occurs via a global rather
than a local instability. Our result is also consistent with previous
simulations of gravitational instability in massive disks by
\citet{laughlin94}, which show that most power goes into the $m=1$
mode.

The behavior of disks in our simulations is significantly different
than that seen in the simulations of \citet{lodato05} and
\citet{rice05}, who model the evolution of disks with masses ranging
from 10\% to 100\% of the primary object mass using polytropic
equations of state with $\gamma>1$, with added cooling terms that
remove energy on time
scales from $3-13\Omega^{-1}$. They find that for all their runs
angular momentum transport is primarily local and that accretion
occurs on a viscous rather than a dynamical time scale, with values of
$\alpha \le 0.06$ in all stable disks. When spiral arms form, the
majority of the power is in $m=2$ modes. Disks settle into steady
states except when $M_d \gtsim M_*$. 

This difference in behavior is likely to be a real physical effect,
caused by two differences between the properties of our disk and those
of \citeauthor{lodato05} and \citeauthor{rice05}
First, the temperature in our disk is set almost entirely by
radiative heating and cooling, in contrast to the polytropic plus
cooling runs in which disk temperatures are set by the balance between
viscous heating and radiative cooling. Viscous heating does not
provide enough energy to raise the disk temperature significantly in
our run.
As a result, the temperature in our disk is almost entirely a
function of distance from the primary star, with no significant
variation at a given distance due to spiral arms or other density or
velocity structures in the disk. Consequently, the thermodynamic
behavior of our disk is closer to an isothermal equation of state than
to the stiffer equations of state produced by values of $\gamma>1$ and
cooling time scales longer than the disk orbital period. This favors
the growth of large-scale global modes that produce rapid angular
momentum transport, an effect pointed out by \citeauthor{lodato05} to
explain the differences between their simulations and those of
\citet{laughlin94}.

Second, in our simulation the disk is never stable or isolated. The
average accretion rate of $4\times 10^{-4}$ $\msun$ yr$^{-1}$ from 17
to 20 kyr, assuming all mass that reaches the star is processed
through the disk, corresponds to $\sim 30\%$ of the disk mass per
orbital period. This is obviously a huge perturbation. Most of this
accretion comes from a large filament (as shown for example in Figure
\ref{100Afull}) that is sheared out into a disk as it approaches the
protostar, an effect that obviously favors $m=1$ spiral
structure. Partly as a result of this perturbation, our disk is never
able to settle into a quasi-steady state, and it forms several
fragments. These likely aid in shepherding material inward into the
primary star. 

These two effects suggest that our results are not
inconsistent with the findings of \citet{lodato05} and \citet{rice05},
simply that our disk is in a different regime of parameter space
than they have explored.

\subsubsection{Disk Fragmentation}
\label{diskfragment}

To help understand why our disk fragments, we compute the
\citet{toomre64} parameter
\begin{equation}
Q \approx \frac{\Omega c_s}{\pi G \Sigma}.
\end{equation}
We do this at each point, and we also compute the mass-weighted
azimuthal average 
\begin{equation}
\left\langle Q\right\rangle_{\phi}(r) \equiv 
\frac{
\int_{-\pi}^{\pi} d\phi \, \Sigma(r,\phi) Q(r,\phi)
}{
\int_{-\pi}^{\pi} d\phi \, \Sigma(r,\phi)
}
\end{equation}
as a function of radius. Obviously this calculation
is somewhat approximate, since we do not have a thin, steady,
Keplerian disk with a well-defined edge as a background
state. Nonetheless, it can give us some insight into the state of the
disk and the reason that it fragments.

\begin{figure}
\plotone{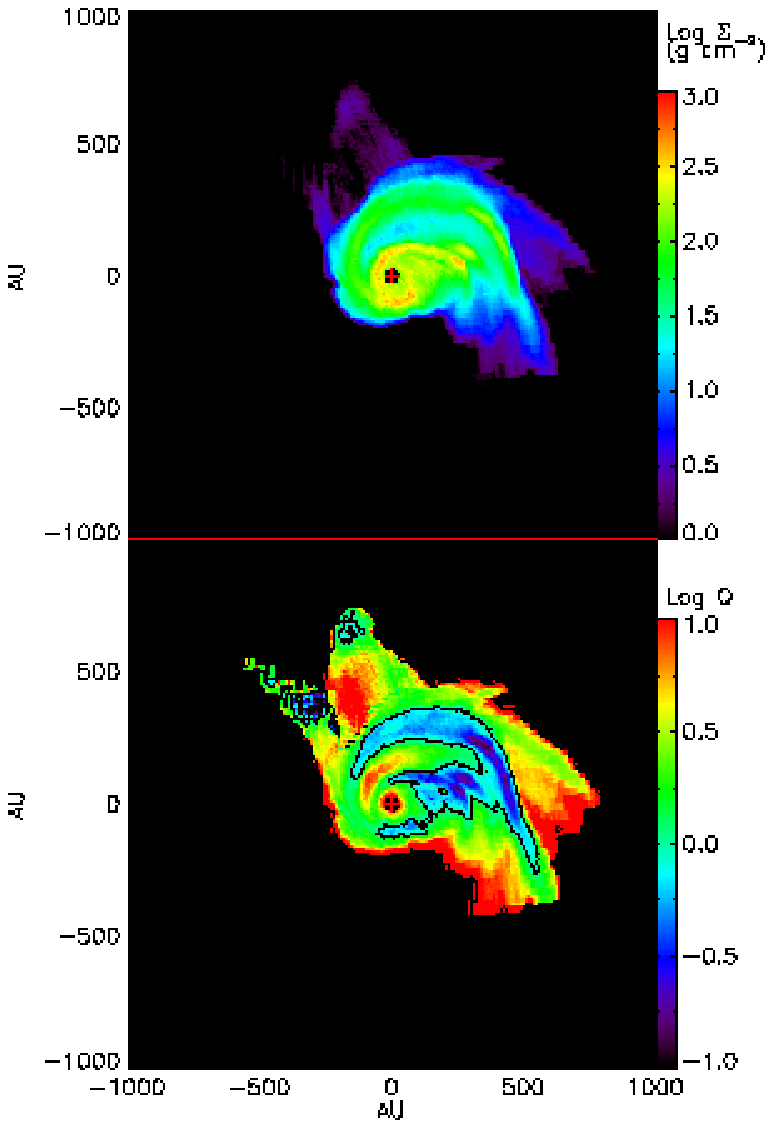}
\caption{
\label{100Adisk1}
Deprojected column density (\textit{upper panel}) and Toomre $Q$
(\textit{lower panel}) of the disk around the primary star in run 100A
at $17.4$ kyr, just as the first disk fragmentation occurs. The
positions of stars are indicated by the white plus signs. In the plot
of $Q$, the black contour indicates $Q=1$, and the exterior black
region consists of points for which there is no gas above the density
threshold we use to define the edge of the disk.
\\
}
\end{figure}

\begin{figure}
\plotone{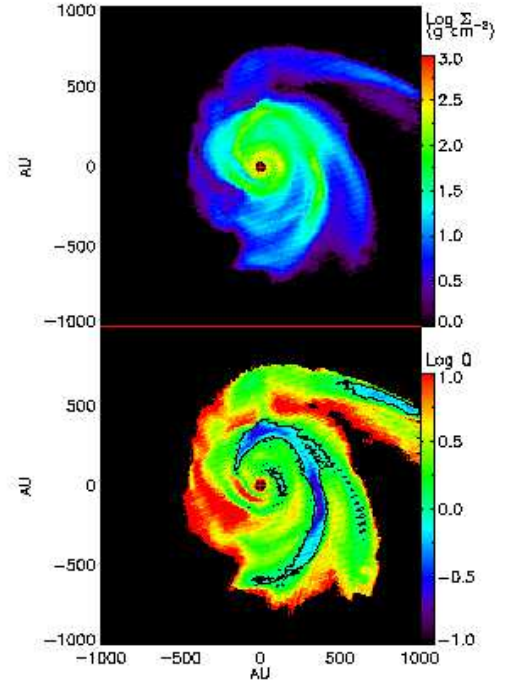}
\caption{
\label{100Adisk2}
Deprojected column density (\textit{upper panel}) and Toomre $Q$
(\textit{lower panel}) of the disk around the primary star in run 100A
at $20$ kyr. For details see Figure \ref{100Adisk1}.
\\
}
\end{figure}

\begin{figure}
\plotone{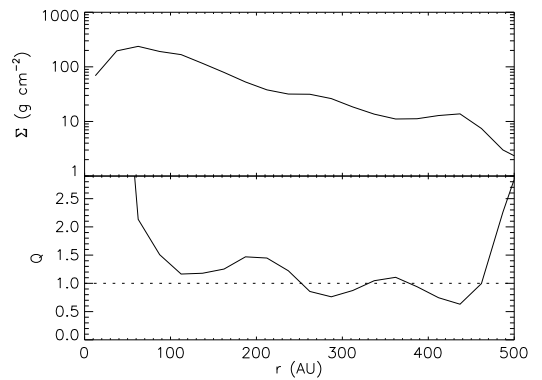}
\caption{
\label{100A1Qaz}
Azimuthally-averaged column density (\textit{upper panel}) and Toomre
$Q$ parameter (\textit{lower panel}) as a function of radius in the
deprojected protostellar disk at $17$ kyr, just before the first disk
fragment forms.
\\
}
\end{figure}

\begin{figure}
\plotone{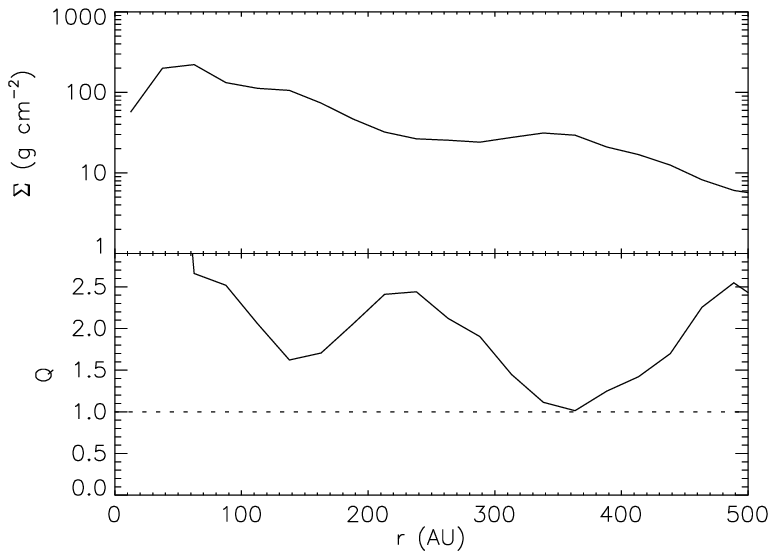}
\caption{
\label{100A2Qaz}
Same as Figure \ref{100A2Qaz}, but at 20 kyr.
\\
}
\end{figure}

Figures \ref{100Adisk1} and \ref{100Adisk2} show plots of deprojected
column density and $Q$ for the disk in run 100A at a time of $17$
kyr, just before the first fragment forms in the disk, and at $20$ kyr,
when the disk-formed star is still present but there are no other
obvious fragments forming. We show the azimuthally-averaged column
density and Toomre $Q$ as a function of radius at these two times in
Figures \ref{100A1Qaz} and \ref{100A2Qaz}. As the plots show, at $17$
kyr there is a broad region where $Q<1$, both at individual points and
in the azimuthal average. This corresponds to the approximate location
where the fragment forms. At 20 kyr, it is clear that
there is another region of the disk that has $Q<1$ at individual
points, but there is no region where the azimuthally-averaged value of
$Q<1$ (although the ring at 350 AU is extremely close). In general,
the disk seems more stable than at 17 kyr, which likely explains why
there is no obvious fragment formation at 20 kyr. However, it seems
entirely possibly that further evolution would decrease $Q$ and lead to
additional fragment formation. Although the luminosity of the star
will continue to rise as it gains mass, heating the disk, the disk
mass will also rise, increasing its shielding against protostellar
radiation. It is not clear which of these effects will dominate.

While fragment formation in massive protostellar disks
is an interesting phenomenon, it seems unlikely to be a significant
hindrance to accretion onto the primary star at this point in the
evolution. There is no noticable drop in the accretion rate onto the
primary after 17 kyr, and from 17 kyr to 20 kyr the primary star mass
increases by a factor of 6 more than the mass of the embedded
fragment. Clearly, most of the mass in the disk is going into the
primary, not into disk-formed fragments.

Our findings on disk fragmentation are broadly consistent with the
analytic predictions of \citet{kratter06}, who analytically model
massive protostellar disks and find that they are unstable to
fragmentation at radii $\gtsim 150$ AU for central stars of mass $M_*
\gtsim 4$ $\msun$. \citeauthor{kratter06}, extending work of
\citet{matzner05}, predict that steady-state disks should fragment if
their sound speeds fall below a critical value $c_{\rm crit} \approx
1.04 (G\dot{M})^{1/3}$, where $\dot{M}$ is the accretion rate onto the
star-disk system. 

Before applying this condition, we modify it in two
ways. First, rather than using the isothermal sound speed to compare
to $c_{\rm crit}$ as \citeauthor{matzner05} suggest, we use
the adiabatic sound speed because our disks are optically thick
to their own radiation. Second, we modify the criterion to account for
the fact that angular momentum transport in our disks appears to be
due to a global rather than a local gravitational
instability. \citeauthor{matzner05} and \citeauthor{kratter06}
determine disk stability using the criterion of \citet{gammie01}, who
simulates angular momentum transport by local gravitational
instabilities and finds that these produce a maximum effective
viscosity $\alpha=0.23$. The critical sound speed depends on $\alpha$
as $c_{\rm crit}\propto \alpha^{1/3}$, because $\alpha$ determines the
rate at which material is processed through the disk onto the central
object. Consequently, the increased rate of angular momentum transport
in our disks makes disks more stable.

With these two modifications, the \citeauthor{kratter06} critical
temperature for instability to fragment formation becomes
\begin{eqnarray}
T_{\rm crit} & \approx & 0.41 \frac{\mu}{\gamma k_B} (\alpha G
\dot{M})^{2/3} \\
& = & 39 \alpha^{2/3}
\left(\frac{\dot{M}}{10^{-4}\,\msun\mbox{yr}^{-1}}\right)^{2/3}
\mbox{ K}
\end{eqnarray}
For the mean accretion rate of $4\times 10^{-4}$ $\msun$ yr$^{-1}$
from 17 kyr to 20 kyr, and our range of estimates $\alpha=1.0-1.6$,
this gives $T_{\rm crit} \approx 70-100$ K. The temperature in the
outer parts of our disks is generally in this range, which explains
why they are marginally unstable to fragment formation. It also
explains why our typical fragmentation radius is somewhat larger than
\citeauthor{kratter06} predict.

\section{Discussion}
\label{discussion}

\subsection{Discussion of Physical Approximations}
\label{limitations}

Our physical formulation of the problem contains three significant
simplifications. Here we discuss them, with the goal of assessing how
much they might affect our results.

First, our treatment of radiative transfer, though a
significant improvement on previous three-dimensional calculations
that ignored radiation entirely, is still quite idealized. Our approach is
gray, so we miss effects that arise from the frequency-dependent
opacities of dust grains. \citet{preibisch95} and
\citet{yorke02}, based on two dimensional calculations, find
that, compared to gray, multi-frequency calculations generally
produce higher dust temperatures and greater degrees of anisotropy in
the radiation field. The fact that, even with the anisotropy effect,
gray radiation almost always underestimates the true temperature
suggests that our results on fragmentation are fairly secure, since
higher temperatures would further reduce the level of
fragmentation. Nonetheless, increasing the anisotropy of the radiation
could could conceivably leave some parts of the flow cooler than we
find, so it would be useful to repeat some of the calculations we
present here with a multi-frequency radiative transfer code.

In addition to being gray, our radiative transfer approach uses the
flux-limited diffusion approximation, which is an approximation that
is only highly accurate in regions that are very optically thick (or
that are very optically thin and for which the geometry is simple). Our
cores have initial mean surface densities of $0.66$ g cm$^{-2}$, which
gives them optical depths $\gtsim 1$ to infrared photons. Thus, the
cores overall are marginally optically thick on average. However, in
the dense regions where fragmentation takes place and where it is most
important to treat the radiation correctly, surface densities are more
typically tens or hundreds of g cm$^{-2}$. Thus, the regions in which
fragmentation occurs are extremely well-described by the
diffusion approximation. Thus, we consider it extremely unlikely that
our results would change qualitatively if we were to use a more
accurate treatment of the radiation field.

A second limitation to our approach is the uncertainty in what our
initial and boundary conditions should be, based on our imperfect
knowledge of the properties of massive cores and their
environments. The MT03 model which we have used fits the
data reasonably well, but observations to date reveal only a little
about the internal structure of massive cores. The primary
uncertainties likely to affect our results are the degree of central
concentration, the amount of internal turbulent structure, and the
nature of any external perturbations. The first of these is difficult
to determine because observations of massive core gas are generally
made with interferometers such as the SMA, which remove large-scale
power and thus make it difficult to determine quantities like
large-scale density gradients. The amount of internal turbulent
structure is poorly known observationally simply because of resolution
and sensitivity limits. Massive cores are too small for observations
to determine fine details of their internal structure. The nature of
external perturbations is uncertain because massive cores are embedded
within clouds that are themselves turbulent, so rather than providing
a constant pressure boundary, the external environment may fluctuate
and drive turbulent motions in a massive core.

One might expect that the stronger the central concentration, the less
fragmentation will occur. However, non-radiative models produce a
great deal of fragmentation regardless of the initial degree of
central concentration, even for cases much more concentrated than the
$k_\rho=1.5$ density gradient we use \citep[e.g.][]{bate03, goodwin04,
dobbs05}. Thus, the degree of initial concentration in our models
seems unlikely to change our results qualitatively. For turbulent
density structure, whether it is structure present initially or
structure induced by external perturbations on a core's surface,
obviously a higher degree of internal structure
favors more fragment formation, and massive cores are likely to have
some internal structure because they are turbulent. Thus, we do not
regard our finding on the absolute number of fragments formed to be
definitive for what will happen in real high-mass cores. However,
since we use identical initial conditions for the radiative and
isothermal runs, the differences between them are likely to remain
even for more structured or less concentrated initial density
fields. Morever, the general effect we have found, that radiative
heating can shut off collapse to small fragments, while allowing the
first object formed to grow to large masses by accreting gas that
cannot otherwise collapse, should apply regardless of the initial and
boundary conditions. Thus, although the quantitative results may vary
for different initial or boundary conditions, the qualitative results
that massive cores do not fragment very strongly should be robust.

The third major limitation to our approach is that we have neglected
magnetic fields. This simplification is partly justified by
observational ignorance. Even in nearby low-mass star-forming regions
there is considerable controversy over how dynamically significant
magnetic fields are \citep[e.g.][]{crutcher99, padoan04b, tassis04},
and observations of more distant, obscured high-mass star forming
regions are far more difficult. \citet{crutcher05} reviews the
available data and concludes that magnetic fields are marginally
dynamically significant, but this conclusion is highly uncertain due
to potential systematic errors in transforming the observed signal
into a magnetic field strength (see \citealt{krumholz06f} for a
discussion of this point). Even if magnetic fields are dynamically
significant in the initial core, simulations show that turbulence can
significantly accelerate the rate of ambipolar diffusion
\citep{heitsch04}, so the field might diffuse out quickly and have
little effect on the overall evolution. Moreover, even if magnetic
fields are dynamically important and can modify how fragmentation
proceeds, our result that radiative transfer suppresses fragmentation
should still hold qualitatively. 

A final magnetic effect that could be
significant is on our protostellar disks. Since we have no magnetic
fields, we obviously have no magnetorotational instability and its
associated angular momentum transport. It is unclear if the MRI can
operate in massive protostellar disks, because their column densities
of $\sim 100$ g cm$^{-2}$ may render them so opaque to protostellar
ultraviolet radiation that their ionization fractions will be too low
for the MRI to operate. (MRI requires sufficient ionization for the
magnetic Reynolds number to be greater than about unity,
\citealt{sano98}.) However, if the MRI does operate, its effect
should be to increase the rate of angular momentum transport, which
will raise the mass of the primary and lower the surface density of
our disks, thereby reducing their propensity to fragment. Thus, if
anything we overestimate fragmentation in our radiative runs.

\subsection{Numerical Resolution}
\label{resolution}

Another potential concern is whether our results depend on our
numerical resolution. At some level, they probably do. The
accretion radius around our sink particles is 4 cells, which
corresponds to $30$ AU in runs 100A, 100B, and 100ISO, and $43$ AU in
run 200A. This means we are unable to resolve binaries whose closest
approach is smaller than twice this, and stars that should become
tight binaries will instead be merged in our code. However, given that
except in the isothermal run, the amount of mass gained by mergers is
negligible, this effect cannot be significant in setting the primary
star mass. We are also
insensitive to the formation of fragments on scales smaller than the
accretion radius, so we could conceivably underpredict the number of
fragments. In our radiative runs this is unlikely to be a problem,
because, as Figure \ref{100Atemp} shows, all the gas
within $\sim 30$ AU of the primary protostar is heated to $\gtsim 300$ K
even at very early times. Thus, heating should very strongly suppress
fragment formation there. On the other hand, this effect could be very
significant in the isothermal run, since isothermal calculations with
higher resolution than we have used do find significant amounts of
fragment formation on $\ltsim 100$ AU scales
\citep[e.g.][]{bate03,dobbs05}, and simulations with varying
resolution find that the amount of fragmentation and the mean fragment
mass are resolution-dependent \citep{martel06}.

A related concern is that we might be missing fragmentation in our
disks due to excessive numerical viscosity. Our Cartesian grid
produces a numerical viscosity that gives an effective $\alpha$ of
\citep{krumholz04}
\begin{eqnarray}
\alpha & \approx &
78 \frac{r_{\rm B}}{\Delta x} \left(\frac{r}{\Delta x}\right)^{-3.85}
\\
& \approx & 0.72 \Delta x_{7.5}^{-2.85} M_1 T_{100}^{-1}
\left(\frac{r}{100\mbox{ AU}}\right)^{-3.85},
\end{eqnarray}
where $r$ is the distance from the star about which the disk orbits,
the mass of the star is $M_1$ in units of $\msun$, the disk
temperature in units of 100 K is $T_{100}$, and $\Delta x_{7.5}$ is
the grid spacing in units of 7.5 AU. This means that, at distances
$\sim 300$ AU, where much of the disk mass resides, our typical
numerical $\alpha$ is of order $10^{-2}$, insignificant compared to
that induced by the SLING instability. On scales $\ltsim 100$ AU,
however, numerical viscosity is probably significant in
shaping the evolution of our disks, and may inhibit fragment
formation. This problem, to the extent that it is significant,
almost certainly affects the isothermal run more than the others,
since irradiated disks at such small radii are quite hot and thus
resistant to fragmentation. The problem may also be more severe in run
200A, where the larger cell size means that the untrustworthy region
is a factor of $2-3$ larger. This may explain the reduced disk
fragmentation we find in that run.

\subsection{Implications for Massive Star Formation}

One of the ouststanding problems in massive star formation has been
how to gather enough mass to make a massive star. Our simulations,
coupled with other recent work, suggest that we may be nearing a
solution. Observations now unambiguously reveal that
there are massive, centrally concentrated cores \citep[e.g. see review
by][]{garay05}. How these cores form is a topic of active research
\citep[e.g.][]{padoan02, tilley04, li04}, but since we can determine
massive core properties from observation we need not solve this
problem to model their evolution.

While it is appealing to see massive cores as the progenitors of
massive stars, one might legitimately worry that these
objects fragment to produce large numbers of low-mass stars rather
than a few massive stars. Some of these low-mass stars could possibly
gain mass via competitive accretion of gas \citep{bonnell01a,
bonnell01b, bonnell04} or via stellar collisions \citep{bonnell98,
bonnell05}, becoming massive stars. However, the former mechanism
seems not to work unless star clusters are typically sub-virial,
globally collapsing objects \citep{krumholz05e, bonnell06c,
krumholz06f}, which appears to be ruled out by observations of the
star formation rate and age spread in young clusters \citep{tan06a,
krumholz06c}. The latter possibility requires stellar densities that
would be very difficult to achieve without global collapse, and which
are orders of magnitude larger than the highest observed stellar
densities in young clusters.

Our results suggest that the solution to this dilemma is that massive
stars do form directly from the collapse of massive cores
(MT03), and that these cores do not fragment strongly because
radiation feedback effectively shuts off fragmentation
\citep{krumholz06b}. In a massive, collapsing core,
most of the mass goes into one primary object. Thus, massive cores --
objects that we know exist from observations -- are the direct
progenitors of massive stars, with no need for intermediate steps of
competitive accretion or collisions.

In this work we have not addressed the question of whether, at higher
masses, radiation pressure might halt accretion and thereby prevent
massive cores from making massive stars. However, theoretical work to
date suggests that radiation beaming by a combination of protostellar
outflow cavities, the protostellar disk, and the accreting envelope
\citep{yorke02, krumholz05d, krumholz05a} provide a robust mechanism
for allowing accretion to continue in the face of radiation
pressure. We plan to report on three dimensional
radiation-hydrodynamic simulations of this problem in future work.

\subsection{Implications for Future Simulations}

Our results show definitively that radiative transfer significantly
modifies the manner in which accretion and fragmentation occur in the
environments where massive stars form, at least on the size scales of
individual cores. The effective heating radius is $>1000$ AU even before
nuclear burning starts in any protostar, simply due to accretion
luminosity. Once nuclear burning starts, the heating radius will rise
rapidly, since the luminosity rises as roughly $M^3$.
It is not possible to capture this effect simply by using a
modified equation of state, because the heating process depends on the
radiative transfer of energy from gas falling onto protostellar
surfaces to gas in the surrounding envelope. We conclude that
simulations of massive star formation and star cluster formation that
do not include radiative transfer and accretion luminosity are not
reliable on size scales below several thousand AU, and that such
calculations almost certainly overestimate the amount of fragmentation
that occurs. This is true even before any of the stars formed begin
nuclear burning. If one wishes to continue a simulation to the point
where deuterium burning starts, one must include that effect as well.

A critical question, which our work raises but does not address, is
the extent to which the overfragmentation problem affects simulations
of low-mass star formation. In such environments cores are usually
separated by more than 1000 AU, accretion rates and the resulting accretion
luminosities are lower, and lower column densities produce lower
opacities to what radiation there is. Thus, we expect the effect on
fragmentation to be less severe than we have found. Nonetheless, it
seems likely that there will be some effect, particularly for models
in which brown dwarfs or low-mass stars form by disk fragmentation
\citep[e.g.][]{bate02a, bate03, goodwin04}, and for competitive
accretion models in which numerous brown dwarfs or low-mass stars form
in clusters $\ltsim 1000$ AU in size, and then evolve in a manner
dictated largely by N-body interactions \citep{bate05, bonnell05}. In
particular, \citet{matzner05} argue that radiative transfer effects
are likely to prevent the formation of brown dwarfs by disk
fragmentation, and our results support the idea that this effect might
be important. It is therefore critical to repeat these calculations
with radiative transfer to see if the fragments persist once better
physics is included, a point also made by \citet{whitehouse06}.

\section{Conclusion}
\label{conclusion}

We report the results of simulations of the collapse and fragmentation
of massive protostellar cores using gravito-radiation hydrodynamics
with adaptive mesh refinement. We find that including radiative
transfer in our simulations produces dramatic effects on the evolution
of these objects. When radiation is included, massive protostellar
cores with the properties of observed cores do not fragment
strongly. They collapse to a handful of objects, with the majority of
the mass accreting onto one primary object. The object gains mass by
accretion of gas that is prevented from collapsing by radiative
heating. Some low-mass stars do form in addition to the primary
massive star, through a combination of fragmentation at sites
sufficiently far from the protostar to be fairly cool, and
fragmentation of unstable protostellar disks around the massive
star. However, these do not contain significant mass. The disks that
form are able to transport mass onto the central star very rapidly due
to a large-scale gravitational instability, which appears to form due
to the SLING mechanism. Overall, our results
are consistent with the turbulent core model of MT02 and MT03,
with the analytic treatment of radiative suppression of fragmentation
by \citet{krumholz06b}, and with the analytic massive protostellar
disk models of \citet{kratter06}.

Our results suggest that massive cores are the direct progenitors of
massive stars, without an intermediate phase of competitive accretion
or stellar collisions. Both of these mechanisms require that a large
number of protostars form out of dense gas in clusters $\sim 1000$ AU
in size or smaller. However, such strong fragmentation seems not to
occur in the environments where massive stars form, because rapid
accretion rapidly raises the gas temperature and prevents nucleation
of small protostars. Instead, most gas in a massive cores accretes
onto the first object to form, and this object prevents other,
comparable mass stars from forming after it.

One additional conclusion from our work is that one must include
radiative transfer in simulations in order to obtain the correct
behavior on small scales. Prescriptions for the equation of state
based on the barotropic approximation or on optically thin heating and
cooling models severely underestimate gas temperatures because they do
not, and cannot, include heating by accretion luminosity onto a central
protostar. The high densities found in massive protostellar cores mean
that this luminosity can reach thousands of $\lsun$ even for $\ltsim 1$
$\msun$ protostars, an effect that cannot be neglected.

\acknowledgements We thank K.~M. Kratter, S.~S.~R. Offner,
C.~D. Matzner, R.~T. Fisher, J.~M. Stone, and J.~C. Tan for helpful
discussions. Support for this work was
provided by NASA through Hubble Fellowship grant \#HSF-HF-01186
awarded by the Space Telescope Science Institute, which is operated by
the Association of Universities for Research in Astronomy, Inc., for
NASA, under contract NAS 5-26555 (MRK); NASA ATP grants NAG 5-12042
and NNG06GH96G (RIK and CFM); the US Department of Energy at the
Lawrence Livermore
National Laboratory under contract W-7405-Eng-48 (RIK); and the NSF
through grants AST-0098365 and AST-0606831 (CFM). This research was
also supported by grants of high performance computing resources from
the Arctic Region Supercomputing Center; the NSF San Diego
Supercomputer Center through NPACI program grant UCB267; the National
Energy Research Scientific Computing Center, which is supported by the
Office of Science of the U.S. Department of Energy under Contract
No. DE-AC03-76SF00098, through ERCAP grant 80325; and the US
Department of Energy at the Lawrence Livermore National Laboratory
under contract W-7405-Eng-48.


\end{document}